\documentclass[aps,prl,twocolumn,superscriptaddress]{revtex4}
\usepackage{amsfonts}
\usepackage{amsmath}
\usepackage{amssymb}
\usepackage{graphicx}
\usepackage{caption}
\usepackage{bm}
\usepackage{color}
\usepackage{mathrsfs}
\usepackage[colorlinks,bookmarks=true,citecolor=blue,linkcolor=red,urlcolor=blue]{hyperref}
\usepackage{appendix}
\usepackage{float}
\usepackage{array}
\usepackage{booktabs}
\newcommand{\RNum}[1]{\uppercase\expandafter{\romannumeral #1\relax}}

\begin{document}
\title{Quasi-symmetry groups and many-body scar dynamics}
\author{Jie Ren}
\thanks{These authors contributed equally to this study.}
\affiliation{Beijing National Laboratory for Condensed Matter Physics and Institute of Physics, Chinese Academy of Sciences, Beijing 100190, China}
\affiliation{University of Chinese Academy of Sciences, Beijing 100049, China}
\author{Chenguang Liang}
\thanks{These authors contributed equally to this study.}
\affiliation{Beijing National Laboratory for Condensed Matter Physics and Institute of Physics, Chinese Academy of Sciences, Beijing 100190, China}
\affiliation{University of Chinese Academy of Sciences, Beijing 100049, China}
\author{Chen Fang}
\email{cfang@iphy.ac.cn}
\affiliation{Beijing National Laboratory for Condensed Matter Physics and Institute of Physics, Chinese Academy of Sciences, Beijing 100190, China}
\affiliation{Songshan Lake Materials Laboratory, Dongguan, Guangdong 523808, China}
\affiliation{Kavli Institute for Theoretical Sciences, Chinese Academy of Sciences, Beijing 100190, China}

\begin{abstract}In quantum systems, a subspace spanned by degenerate eigenvectors of the Hamiltonian may have higher symmetries than those of the Hamiltonian itself. When this enhanced-symmetry group can be generated from local operators, we call it a quasi-symmetry group. When the group is a Lie group, an external field coupled to certain generators of the quasi-symmetry group lifts the degeneracy, and results in exactly periodic dynamics within the degenerate subspace, namely the many-body-scar dynamics (given that Hamiltonian is non-integrable). We provide two related schemes for constructing one-dimensional spin models having on-demand quasi-symmetry groups, with exact periodic evolution of a pre-chosen product or matrix-product state under external fields.
\end{abstract}
\maketitle

\paragraph{Introduction}
Symmetry plays a central role in physics.
Given a quantum system described by Hamiltonian operator $\hat{H}$, a symmetry $g$, restricted to be unitary in this work, is represented by a unitary operator $\hat{D}(g)$ such that
\begin{equation}\label{eq:1}
[\hat{H},\hat{D}(g)]=0.
\end{equation}
If multiple $g$'s form a group $G$, Eq.(\ref{eq:1}) leads to the fundamental theorem that each eigen-subspace $\Psi_E \equiv \{\psi|\hat{H}\psi=E\psi\}$ is invariant under $\hat{D}(g)$ for $g\in{G}$, or one can casually say that $\Psi_E$ at least has symmetry group $G$.
In other words, generally, $\Psi_E$ has \emph{higher} symmetry than $G$.

As an example, consider two $1/2$-spins coupled by a Heisenberg interaction, $\hat{H}=\mathbf{\hat{S}}_1\cdot\mathbf{\hat{S}}_2$.
The full symmetry group of the triplet eigen-subspace is U(3), of which the Hamiltonian symmetry group SO(3) is a subgroup.
However, not all symmetries in U(3) are physically interesting, because many of them involve creating (annihilating) entanglement between the spins, and as such are difficult to realize in experiments.
Therefore, hereafter we restrict to more physically relevant cases: an operator $\hat{D}(\tilde{g})$ that preserves an eigen-subspace of $\hat{H}$ is considered as a ``symmetry'', if and only if $\hat{D}(\tilde{g})$ is a direct product of unitary operators on individual spins, that is,
\begin{equation}\label{eq:2}
\hat{D}(\tilde{g})=\hat{d}_1(\tilde{g})\otimes\hat{d}_2(\tilde{g})\otimes...\otimes\hat{d}_N(\tilde{g}),
\end{equation}
known as the onsite-unitary condition.
This requires the representation of $G$ to be a tensor-product representation, that is, neither spatial nor time-reversal symmetry is considered, unless otherwise specified.
In the above two-spin example, a unitary operation sending $\left|\uparrow\uparrow \right\rangle$ to $(\left|\uparrow\downarrow\right\rangle+\left|\downarrow\uparrow\right\rangle)/\sqrt{2}$ leaves the triplet eigen-subspace invariant, but cannot decompose as in Eq.(\ref{eq:2}).
In fact, it can be checked that all the symmetries of the triplet eigen-subspace meeting the onsite-unitary condition Eq.(\ref{eq:2}) are just the overall rotations SO(3).
The triplet eigen-subspace has hence the same symmetry group as $\hat{H}$ itself.

The above discussion motivates us to define a new type of symmetry operation, which we tentatively term \emph{quasi-symmetry}, as a unitary operator $\hat{D}(\tilde{g})$ satisfying Eq.(\ref{eq:2}), so that a given eigen-subspace of $\hat{H}$ having energy $E$ is invariant under $\hat{D}(\tilde{g})$.
It is obvious that ${\tilde{g}}$'s as such form a new group, denoted by $\tilde{G}_E$.
We call $\tilde{G}_E$ the quasi-symmetry group of $\hat{H}$ with respect to the eigen-subspace $\Psi_E$.
If $\hat{D}(\tilde{g})$ commutes with $\hat{H}$, then $\tilde{g}$ is a quasi-symmetry for any eigen-subspace of $\hat{H}$, so the symmetry group is always a subgroup of any quasi-symmetry group for a given Hamiltonian: $G\subset\tilde{G}_E$.

Before showing an explicit example of quasi-symmetry in quantum models, we point out that its classical counterpart, known as non-symmetry-caused degeneracy, is well-known in models for frustrated magnetism.
Consider a classical $J_1$-$J_2$-model on a square lattice, where Heisenberg $J_1$ couplings connect nearest spins, and $J_2$ next-nearest-neighbor spins of length $s$.
This Hamiltonian is invariant under any overall SO(3) rotation, but is \emph{not} invariant under relative rotations between the two sublattices.
Nevertheless, consider a state where all spins in each sublattice are antiferromagnetically aligned, then it is easy to check that the energy, being $-2J_2s^2$ per spin, is independent of the relative angle between the two sublattices.
Therefore, a relative rotation between the sublattices, not being a symmetry of $H$, does lead to classical degeneracy.
Can we obtain a quasi-symmetry model by quantizing the above $J_1$-$J_2$-model?
The answer is negative: when quantum fluctuation is turned on, the above classical degeneracy is lifted due to the famous order-by-disorder mechanism \cite{Chandra1990}.

We do not know a deterministic way for diagnosing all possible quasi-symmetries in a given Hamiltonian, quantum or classical.
Yet fortunately, recent progress in the study on quantum-many-body scars \cite{Heller1984,Bernien2017,Turner2017,Turner2018,Moudgalya2018,Moudgalya2018a,lin2018,Khemani2018} provides with many examples of quasi-symmetry in quantum models 
\footnote{Quasi-symmetry without extensive degeneracy in quantum models has appeared ealier in literature. See \cite{SRIRAMSHASTRY} for example.}.
In certain non-integrable quantum many-body systems, there exist some close trajectories in the Hilbert space, along which a special short-range-entangled state evolves periodically or quasi-periodically, independent of the size of the system \cite{Choi2018,Ho2019,Michailidis2019,Bull2019,Hudomal2019,Alhambra2020,turner2020}.
The evolution of certain many-body states along these closed trajectories, as opposed to the chaotic trajectories for generic states, is called the quantum-many-body scar dynamics, or simply scar dynamics.
All the states along one such trajectory span a Hilbert subspace invariant under the Hamiltonian evolution, and the eigenstates of $\hat{H}$ within this subspace form a tower of states, namely the scar tower \cite{Mark2020,Moudgalya2020,Bull2020}.
The scar dynamics is related to the violation of the eigenstate-thermalization hypothesis \cite{Deutsch1991,Srednicki1994,rigol2008,nandkishore2015,dAlessio2016} in certain eigenstates from the scar tower.
In previously studied exact cases \cite{Schecter2019,Iadecola2020,Chattopadhyay2020,Shibata2020,Mark2020a,Moudgalya2020a,Pakrouski2020}, a scar Hamiltonian consists in two parts
\begin{equation}\label{eq:3}
\hat{H}_{scar}=\hat{H}+\hat{H}_1,
\end{equation}
where $\hat{H}$ has a degenerate eigen-subspace $\Psi_E$ and $\hat{H}_1$ (i) preserves the subspace $\Psi_E$ but (ii) lifts the degeneracy by breaking energy spectrum into a ``tower'' with equal spacing $\delta{E}$.
It then becomes obvious that a random initial state in $\Psi_E$ oscillates with a period $2\pi\delta{E}^{-1}$.
If a scar Hamiltonian in Eq.(\ref{eq:3}) satisfies (i) $\hat{H}_1$ is a sum of local operators and (ii) there is at least one product state $\psi_0\in\Psi_E$, then the quantum Hamiltonian $\hat{H}$ has at least $\tilde{G}=\mathrm{U(1)}$ quasi-symmetry $\hat{D}(\tilde{g}(\theta))\equiv\exp(i\hat{H}_1\theta)$ with respect to $\Psi_E$.
In other words, under the above conditions, quantum-many-body-scar dynamics is a sufficient condition for the existence of quasi-symmetry.

Does quasi-symmetry also imply scar dynamics?
Suppose there is a quasi-symmetry group $\tilde{G}_E\neq{G}$ for some $\hat{H}$ with respect to $\Psi_E$.
If $\tilde{G}_E$ is a compact Lie group, then thanks to the onsite-unitary condition Eq.(\ref{eq:2}), we have that any generator
\begin{equation}
\hat{X}=\hat{x}_1\oplus\hat{x}_2\oplus\dots\oplus\hat{x}_N
\end{equation}
is a sum of local operators $\hat{x}_i$'s, each of which is a hermitian operator acting on the $i$-th spin.
Choose $\hat{H}_1=c\hat{X}$ for the scar Hamiltonian in Eq.(\ref{eq:3}), where $c$ is a real constant.
For any state $\psi(t=0)\in\Psi_E$ as initial state, we have
\begin{equation}
\hat{H}\psi(t)=\hat{H}\exp[-i(\hat{H}+\hat{H}_1)t]\psi(t=0)=E\psi(t),
\end{equation}
meaning that $\Psi_E$ is preserved by the scar Hamiltonian $\hat{H}_{scar}$.
Further, if $X$ generates a U(1) subgroup of $\tilde{G}$, then the spectrum of $\hat{X}$ has equal spacing $\Delta$, and the evolution of any $\psi\in\Psi_E$ has exact period $2\pi(c\Delta)^{-1}$.
Therefore, quasi-symmetry Lie group in $\hat{H}$ indeed implies scar dynamics, given that $\tilde{G}_E\neq{G}$.
When $\tilde{G}_E$ is a discrete group, there is not an obvious choice for a scar Hamiltonian.
In that case, there is a discrete version of scar dynamics, to be discussed in Ref.\onlinecite{SOM}.

In this work, we focus on constructing spin Hamiltonians $\hat{H}$ that have a quasi-symmetry group $\tilde{G}$ of choice.
In the main text, we assume that the quasi-symmetry group be a compact Lie group.
Our construction scheme uses three elements as input: a spin-$s$ spin chain defining the Hilbert space, $s=1/2,1,3/2,\dots$, a compact Lie-group $\tilde{G}$ of choice, and an ``anchor state'', denoted by $\psi_0$, which is either a product or a matrix-product state \cite{Verstraete2008}.
For simplicity, we in this work only use two anchor states as examples: an all-up ferromagnetic state and an Affleck-Kennedy-Lieb-Tasaki-like \cite{Affleck1987} matrix-product state.
The constructed Hamiltonian $\hat{H}$ is expressed in terms of projectors acting on small clusters, the same as in Ref.\onlinecite{Shiraishi2017}, but the method for defining the small-cluster projectors are based on two inputs: the anchor state and the quasi-symmetry group
\footnote{There are other alternative ways to do the same job, for example, the inverse method in \cite{inversemethod}, but we choose the method that is least technical for better presentation.}.

\paragraph{Product states as anchor states}
\label{sec:PSanchor}
We first describe the construction of spin-$s$ Hamiltonians with a chosen $\tilde{G}$ using the all-up state $\psi_0=|s\dots{s}\rangle$ as the anchor state.
To start with, we consider a cluster of $m$ spins, or simply, an $m$-cluster. The product state $\psi_0$ restricted to an m-cluster is denoted by $\psi_0^{[m]}$.
The unitary operators on a single spin form the unitary group $\mathrm{U(2s+1)}$, and we assume that $\tilde{G}\subset{U}(2s+1)$.
Define $\Psi_{\tilde{G}}^{[m]}$ as the following subspace in the $m$-cluster space
\begin{equation}
\Psi_{\tilde{G}}^{[m]} \equiv span\{\hat{d}^{\otimes{m}}(\tilde{g})\psi_0^{[m]}|\tilde{g}\in{\tilde{G}}\},
\end{equation}
and define $\hat{P}$ as the projector onto $\Psi_{\tilde G}^{[m]}$.
Then we consider the following $m$-cluster Hamiltonian
\begin{equation}\label{eq:7}
\hat{H}_{[m]}=(1-\hat{P})\hat{h}(1-\hat{P}),
\end{equation}
where $\hat{h}$ is an arbitrary hermitian matrix acting on the $m$-cluster.
It is easy to see that $\Psi_{\tilde G}^{[m]}$ is the zero energy subspace of $\hat{H}_{[m]}$ for a randomly chosen $\hat{h}$.

Now, consider an infinite chain.
For each $m$-cluster of consecutive spins we define a term as in Eq.(\ref{eq:7}), and obtain the full Hamiltonian
\begin{equation}\label{eq:8}
\hat{H}=\sum_{j=1,...,N}(1-\hat{P}_{[j,j+m-1]})\hat{h}_{[j,j+m-1]}(1-\hat{P}_{[j,j+m-1]}),
\end{equation}
where $\hat{P}_{[j,j+m-1]}$ is the $m$-cluster projector in Eq.(\ref{eq:7}) over the $j,j+1,\dots,j+m-1$ spins, and $\hat{h}_{[j,j+m-1]}$ is a random hermitian operator on the same cluster.
The summation in Eq.(\ref{eq:8}) is from $j=1$ to $j=N-m+1$ if the chain is open, and to $j=N$ if closed.
Periodic cycling is understood for a closed chain: when $j+l>N$, replace $j+l$ with $j+l-N$.
Two observations can be made: (i) the all-up state $\psi_0$ is a zero-energy eigenstate of $\hat{H}$, because $(1-\hat{P}_{[j,j+m-1]})\psi_0=0$ for each $j$, and (ii) states of the following form
\begin{equation}\label{eq:9}
\hat{D}(\tilde{g})\psi_0\equiv\hat{d}^{\otimes{N}}(\tilde{g})\psi_0,
\end{equation}
are also zero-energy eigenstate of $\hat{H}$ for the same reason.
All $\hat{D}(\tilde{g})\psi_0$'s in Eq.(\ref{eq:9}) and their linear combinations form a subspace $\Psi_{\tilde{G}} \equiv span\{\hat{D}(\tilde g)\psi_0|\tilde g\in \tilde G\}$.
It is clear that $\Psi_{\tilde{G}}\subset\Psi_0$, the zero energy subspace of $\hat{H}$.
The Hamiltonian $\hat{H}$ hence has quasi-symmetry group $\tilde{G}$ with respect to $\Psi_{\tilde{G}}$.

To better illustrate the scheme, we look at one example where $s=1$, $m=2$ and $\tilde{G}=\mathrm{SO(3)}\subset\mathrm{U(3)}$.
For the $2$-cluster, namely the $j$-th spin and the $(j+1)$-th spin, the total spin $S=0,1,2$, and the all-up state $\psi_0^{[2]}=|++\rangle$ belongs to $S=2$-subspace.
Therefore acting $\hat{d}(\tilde{g})\otimes\hat{d}(\tilde{g})$ where $\tilde{g}\in\mathrm{SO(3)}$ on $\psi_0^{[2]}$ yields the entire $S=2$-subspace, which is $\Psi_{\tilde{G}}^{[2]}$.
The 2-cluster projector onto $\Psi_{\tilde{G}}^{[2]}$ is
\begin{equation}
\hat{P}_{[j,j+1]}=(\hat{\mathbf{S}}_j+\hat{\mathbf{S}}_{j+1})^2[(\hat{\mathbf{S}}_j+\hat{\mathbf{S}}_{j+1})^2-2]/24.
\end{equation}
Substituting $\hat{P}_{[j,j+1]}$ and a random choice for $\hat{h}_{[j,j+1]}$ into Eq.(\ref{eq:8}), we have the full Hamiltonian.
An exact diagonalization of this Hamiltonian (with periodic boundary) is carried out for $2\le{N}\le10$.
We plot the level statistics in Ref.\onlinecite{SOM}, which fits the Wigner-Dyson curve, indicating non-integrability of the Hamiltonian \cite{levelstat}.
The diagonalization also shows that there are exactly $2N+1$ independent states in $\Psi_0$, which are nothing but the states in the largest total spin sector (total spin being $N$), and that $\Psi_{\tilde{G}}=\Psi_0$.

We can also choose $\tilde{G}=\mathrm{SU(2)}\subset\mathrm{U(3)}$, and the same $\psi_0$ as the anchor state.
In Ref.\onlinecite{SOM}, we show that the resultant $\Psi_{\tilde{G}}$ (which again equals $\Psi_0$) is exactly spanned by, up to an onsite-unitary transform, the type-I scar tower of the spin-1-XY model in Ref.\onlinecite{Schecter2019}, although the Hamiltonian, due to the randomness in $\hat{h}_{[j,j+1]}$, can be drastically different from that of the XY-model.
(There are two scar towers discovered in Ref.\onlinecite{Schecter2019}, and we denote them, after their sequential appearances in the original paper, as ``type-I'' and ``type-II''. Also see Ref.\onlinecite{Chattopadhyay2020} for more on the type-II case.)

This simple example of SO(3) quasi-symmetry group illustrates some general features of quasi-symmetry groups.
First, $\tilde{G}$ is a subgroup of $\mathrm{U(2s+1)}$, so that by choosing a large $s$ one can specify any compact Lie group, such as SO(n), U(n), Sp(n), and exceptional Lie groups, as the quasi-symmetry group.
We note here that the actual form of the ``sandwiched'' part of the Hamiltonian in Eq.(\ref{eq:8}), $\hat{h}_i$, is almost completely irrelevant, as long as it does not have so many symmetries that the Hamiltonian becomes integrable.
Last, we want to emphasize that, despite the randomness in $\hat{h}_{[j,j+m-1]}$, it is \emph{not} guaranteed that $\Psi_{\tilde{G}}=\Psi_0$.
This indicates that the zero-energy subspace of $\hat{H}$, despite being designed to be so, is not generated by acting $\hat{D}(\tilde{G})$ on $\psi_0$.
This equality between the two can only be established, or disproved, in numerics up to some $N$, as we do in Ref.\onlinecite{SOM}.

\paragraph{Matrix-product states as anchor states}
\label{sec:MPSanchor}

A product state has zero entanglement, and if chosen as the anchor state, or equivalently, the initial state, during the time evolution the state remains a product state, because quasi-symmetry operations are strictly local.
It is natural that we extend the discussion to the case where the anchor state has finite entanglement, i.e., is a matrix-product state.
The corresponding construction of the scar Hamiltonian follows a slightly more complicated scheme, compared with the product-state case.
Again considering a group $\tilde{G}\subset{U(2s+1)}$, we first obtain two linear or projective representations of $V$ of equal dimension $\chi$, $d_{L}(\tilde{G}),d_{R}(\tilde{G})$, such that $d_L\otimes{d_R}$ contains a representation of dimension $2s+1$, denoted by $d(\tilde{G})$.
In other words, there exists a trio of representations $d_{L},d_{R},d$ of dimensions $\chi$, $\chi$ and $2s+1$, such that the Clebsch-Gordon coefficients $\langle{d}_L,\alpha;d_R,\beta|d,k\rangle\neq0$, where $\alpha,\beta=1,\dots,\chi$ and $k=1,\dots,2s+1$.
When these conditions are met, define matrices
\begin{equation}\label{eq:11}
A^{k}_{\alpha\beta}\equiv\langle{d}_L,\alpha;d_R,\beta|d,k\rangle.
\end{equation}
These matrices define our anchor state(s), which is
\begin{eqnarray}
\psi_0&=&Tr(A^{s_1}\dots{A}^{s_N})|s_1,\dots,s_N\rangle,\\
\nonumber
\psi_{\alpha\beta}&=&(A^{s_1}\dots{A}^{s_N})_{\alpha\beta}|s_1,\dots,s_N\rangle
\end{eqnarray}
for a closed and an open chain, respectively.

\begin{figure}[H]
\begin{centering}
\includegraphics[width=.9\linewidth]{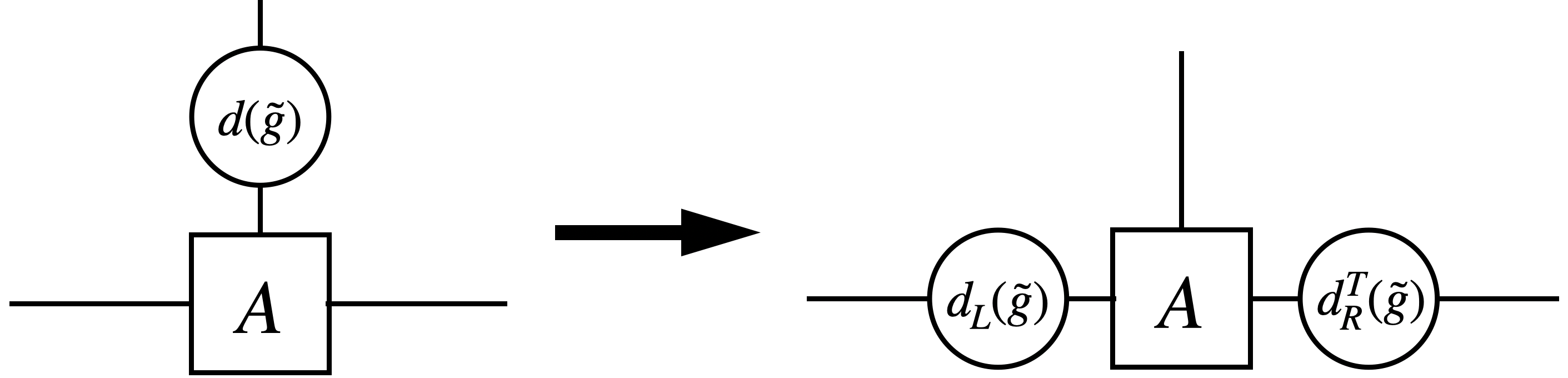}
\par\end{centering}
\caption{Action of onsite operator $\hat{d}(\tilde{g})$ on the Clebsch-Gordon coefficients tensor $A$, the representation $d(\tilde G)$ on the physical indices is transfered to two (projective) representation $d_{L,R}(\tilde G)$ on the bond indices.}
\end{figure}

Consider an $m$-cluster, on which the matrices Eq.(\ref{eq:11}) define $\chi^2$ open-matrix-product states
\begin{equation}
\psi_{\alpha\beta}^{[m]}=(A^{s_1}\dots{A}^{s_m})_{\alpha\beta}|s_1,\dots,s_m\rangle,
\end{equation}
where $\alpha,\beta=1,\dots,\chi$.
Acting $\hat{d}^{\otimes{m}}(\tilde{g})$ for any $\tilde{g}\in{\tilde{G}}$ on these $\chi^2$ states yields another set of $\chi^2$ open-matrix-product states:
\begin{eqnarray}
\nonumber
&&\langle{s}_1\dots{s}_m|\hat{d}^{\otimes m}(\tilde{g})\psi_{\alpha\beta}^{[m]}\rangle \\ 
&\equiv & {d}_{s_1s'_1}(\tilde{g})\dots{d}_{s_ms'_m}(\tilde{g})[A^{s'_1}\dots{A}^{s'_m}]_{\alpha\beta} \nonumber \\
&=&[d_L(\tilde g)A^{s_1}{d}_R^T(\tilde{g})\dots d_L(\tilde g){A}^{s_m} d^T_R(\tilde g)]_{\alpha\beta}.
\end{eqnarray}
Find the subspace
\begin{equation}
\Psi_{\tilde{G}}^{[m]} \equiv span\{\hat{d}^{\otimes m}(\tilde g)\psi_{\alpha\beta}^{[m]}|\tilde g \in \tilde G,\alpha,\beta=1,\cdots,\chi\},
\end{equation}
and define $\hat{P}$ as the projector onto to $\Psi_{\tilde G}^{[m]}$.

For a closed chain of $N\ge{m}$ sites, define the Hamiltonian as in Eq.(\ref{eq:8}).
It is easy to verify that the anchor state $\psi_0$ is a zero eigenstate of $\hat{H}$ because it is a zero eigenstate of each term;
and also the state $\hat{D}(\tilde g)\psi_0\equiv\hat{d}^{\otimes{N}}(\tilde g)\psi_0$ is a zero eigenstate for the same reason for $\tilde g \in \tilde G$.
The space $\Psi_{\tilde G}$ spanned by all these states is thus a zero-energy subspace of $\hat{H}$, i.e., $\Psi_{\tilde G}\subset\Psi_0$.
Therefore, we have constructed $\hat{H}$ that has quasi-symmetry group $\tilde{G}$ with respect to $\Psi_{\tilde G}$.
The case of open chains can be similarly worked out (not shown here).

We again use an example to illustrate the above construction scheme.
Choose $\tilde G=\mathrm{U(1)}\subset\mathrm{U(3)}$ as our quasi-group, and we choose $d_L(\tilde G)=d_R(\tilde G)=\frac{1}{2}\oplus-\frac{1}{2}$ that are the two-dimensional reducible projective representations of U(1).
The specific realization of U(1) can be arbitrary, but in this example we choose it to be the overall spin rotation about $z$-axis.
$d(\tilde G)$ is chosen to be the three-dimensional reducible vector representation $d(\tilde G)=(x,y,z)=+1\oplus0\oplus-1$.
So the matrices are given by the Clebsch-Gordon coefficients
\begin{equation}\label{eq:MPS}
A^{\pm}=\sqrt{\frac{1}{6}}(\sigma_0\pm\sigma_z),A^0=\sqrt{\frac{1}{3}}\sigma_x
\end{equation}
satisfying
\begin{equation}
\exp(i\hat{S}_{z}\theta)_{ij}A^j=e^{i\sigma_z\theta/2}A^i(e^{i\sigma_z\theta/2})^T.
\end{equation}
Now we consider an $m=3$-cluster.
The four open $3$-cluster states, $\psi_{\alpha\beta}$, are none but the Affleck-Kennedy-Lieb-Tasaki open $3$-chain ground states, up to a unitary transform $\exp(iS_y\pi)$ on all odd sites.

After acting all elements of the U(1) quasi-symmetry group on the four open $3$-cluster states, we have a subspace $\Psi_{\tilde G}^{[3]}$ spanned by 12 states, classified into groups labeled by two quantum numbers $n_\pm\equiv\hat{S}_{1z}\pm\hat{S}_{2z}+S_{3z}$:
\begin{eqnarray}\nonumber
(n_+,n_-)=(0,0): &\frac{|+0-\rangle-|-0+\rangle}{\sqrt{2}}, \frac{|+0-\rangle+|-0+\rangle+|000\rangle}{\sqrt{3}},\\
\nonumber
(\pm1,\pm1): &\frac{|\pm00\rangle+|00\pm\rangle}{\sqrt{2}},\\
\nonumber
(\pm1,\mp1): &|0\pm0\rangle,\\
\nonumber
(\pm2,0): &|\pm\pm0\rangle, |0\pm\pm\rangle,\\
(\pm3,\pm1): &|\pm\pm\pm\rangle.
\end{eqnarray}
Define $\hat{P}$ as the $3$-cluster projector onto $\Psi_{\tilde G}^{[3]}$.
Replacing $\hat{P}_{[j,j+2]}$ with $\hat{P}$ in Eq.(\ref{eq:8}), we have the full Hamiltonian $\hat{H}$ with quasi-symmetry $\mathrm{U(1)}$, with respect to the zero energy subspace $\Psi_{\tilde{G}}$.
Using numerical calculation up to $N=10$ sites \cite{SOM}, we find that level-spacing statistics of $\hat{H}$ shows Wigner Dyson behavior.
We have also checked, up to $N=14$, that the degeneracy of the zero subspace of $\hat{H}$ is $N+1$ for periodic chains and $4N$ for open chains, and that $\Psi_0=\Psi_{\tilde{G}}$.
This means that the entire zero-energy subspace of $\hat{H}$ can be obtained from acting the quasi-symmetry group elements on the anchor state(s).
It is interesting to notice that, after an onsite-unitary transform, the resultant zero-energy subspace becomes the space spanned by the type-II-spin-1-XY scar \cite{Schecter2019,Chattopadhyay2020}.
We comment that since the quasi-symmetry group is only U(1), instead of SU(2) or higher Lie groups, there is not an obvious choice for a local $\hat{Q}$ such that $[\hat{Q},\hat{H}]=const*\hat{Q}$ on the subspace.
We also remark that the Hamiltonian following our construction is ``unfrustrated'', in the sense that $\Psi_0$ lies within the zero-energy subspace of each term in $\hat{H}$, in contrast to the original XY-model.
It is certainly possible to construct models having larger quasi-symmetry groups, such as $\mathrm{SO(3)}$, using the same MPS as in Eq.(\ref{eq:MPS}), an explicit example of which is shown in Ref.\onlinecite{SOM}.

We comment that using matrix-product states as anchor states is particularly useful when we relate this study to the study of symmetry-protected topological states \cite{Gu2009,Chen2012,Lu2012} (SPT).
In Ref.\onlinecite{SOM}, we show how one can construct a scar tower and Hamiltonian such that all states of the form $\hat{D}(\tilde{g})\psi_0$ is an SPT protected by a unitary or anti-unitary group.
Here we simply point out that in the example above, both $\psi_0$ and $\hat{D}(\tilde{g})\psi_0$ are SPT protected by time-reversal symmetry, demonstrated in Ref.\onlinecite{SOM}.

\paragraph{Discussion}
\label{sec:Dis}
Aiming for a simple narrative, we have so far assumed that the anchor states have translation symmetry, and the quasi-group symmetry operator $\hat{D}(\tilde{g})$ acts uniformly on each spin, as in Eq.(\ref{eq:9}).
Both conditions can be relaxed: (i) the anchor state may be rotated by onsite-unitary operators $\hat{d}_1(\tilde{g}_1)\otimes\hat{d}_2(\tilde{g}_2)\otimes\dots\otimes\hat{d}_N(\tilde{g}_N)$ for $\tilde{g}_i \in \tilde{G}$; (ii) the action of $\hat{D}(\tilde{G})$ can be generalized to
\begin{equation}
\hat{D}(\tilde{g})=\hat{d}_1(\tilde{g})\otimes\hat{d}_2(\tilde{g})\otimes\dots\otimes\hat{d}_N(\tilde{g}),
\end{equation}
where $\hat{d}_{i=1,\dots,N}$ are $N$ different representations of $\tilde{G}$.
With these generalizations, the method for defining the $m$-cluster projectors becomes slightly modified, shown in Ref.\onlinecite{SOM}.

The anchor state, product or matrix-product, is a key input for our construction scheme.
It ensures that within the zero-energy subspace of constructed Hamiltonian, there is at least one state that is a (matrix) product state.
The anchor state can also be used as the initial state in the associated scar dynamics, and due to the onsite-unitary condition, all the states along the entire trajectory are (matrix) product states as the anchor state.
In previous studies, the state used as the origin, from which the scar tower is obtained using ladder operators, is an exact eigenstate of the scar Hamiltonian, rather than an initial state for scar dynamics.

We impose the quasi-symmetry group $\tilde{G}$ without requiring a ladder operator $\hat{Q}$.
However, if $\tilde{G}$ is a non-Abelian Lie group, a ladder operator can always be found, because in that case SO(3)$\subset\tilde{G}$, and SO(3) has ladder operator $\hat{Q}=\hat{L}_x-i\hat{L}_y$.
For $\tilde{G}=\mathrm{U(1)}$, we have used one above example to show that even in the absence of $Q$, the zero-energy subspace of $\hat{H}$ forms a scar tower identical to the type-II-spin-1-XY scar tower.
On the other hand, if $\tilde{G}\supset{SO(3)}$, there are in general multiple ladder operators.
For example, when $\tilde{G}=\mathrm{SU(3)}\supset\mathrm{SU(2)}$, there are three different ladder operators, corresponding to the three natural embeddings of SU(2) in SU(3).
See Ref.\onlinecite{SOM} for an explicit model, and a general discussion on the relation between the non-Abelian quasi-symmetry group and ladder operators.

To summarize, we show that many-body-scar towers have hidden group structures that we call quasi-symmetry groups, and propose schemes for constructing local Hamiltonians that host any chosen Lie group as its quasi-symmetry group.
As application of the new concept, we show that (i) several known scar models can be unified; (ii) a scar model having three sets of ladder operators can be found (see Ref.\onlinecite{SOM}); and (iii) a discrete version of many-body-scar is established by choosing a discrete quasi-symmetry group (see Ref.\onlinecite{SOM}).

\begin{acknowledgments}
CF thanks B. A. Bernevig for discussion and feedback, and thanks Huan He for the first introduction to the study of quantum many-body scars.
CL and JR thank Shu Chen for his encouragement and support at the early stage of this project.
CF and JR acknowledges support from Ministry of Science and Technology of China under grant number 2016YFA0302400, National Science Foundation of China under grant number 11674370, and Chinese Academy of Sciences under grant number XXH13506-202 and XDB33000000.
CL acknowledges support from Ministry of Science and Technology of China under grant number 2016YFA0300600.
\end{acknowledgments}

\bibliography{ref}

\onecolumngrid
\newpage

\begin{appendix}
\section{Appendix A: Notations}
In this Appendix, we briefly summarize the notations used in the main text and the supplemental material. The symbols frequently used are listed in the following tables:

\begin{table}[H]
\caption{Notations for (quasi-)symmetry groups}

\centering{}
\setlength{\tabcolsep}{3mm}
\begin{tabular}{cl}
\hline
\hline 
Symbols & Name meaning \tabularnewline
\hline 
\specialrule{0em}{0pt}{1pt}
$G$ & Symmetry group. 
\tabularnewline \specialrule{0em}{1pt}{1pt}
$\tilde G_E$ & Quasi-symmetry group of eigen-subspace with energy $E$. 
\tabularnewline \specialrule{0em}{1pt}{1pt}
$\tilde G$ & Quasi-symmetry group (with energy $E=0$ by default). 
\tabularnewline \specialrule{0em}{1pt}{1pt}
$g$ & Element of symmetry group. 
\tabularnewline \specialrule{0em}{1pt}{1pt}
$\tilde g$ & Element of quasi-symmetry group. 
\tabularnewline \specialrule{0em}{1pt}{1pt}
$\hat d_i(g)$ & Single-site representation operator(matrix) of group element $g$ in $G$. 
\tabularnewline \specialrule{0em}{1pt}{1pt}
$\hat d_i(\tilde g)$ & Single-site representation operator(matrix) of group element $\tilde g$ in $\tilde G$. 
\tabularnewline \specialrule{0em}{1pt}{1pt}
$\hat D(g)$ & Many-body representation operator(matrix) of group element $g$ in $G$. 
\tabularnewline \specialrule{0em}{1pt}{1pt}
$\hat D(\tilde g)$ & Many-body representation operator(matrix) of group element $\tilde g$ in $\tilde G$. 
\tabularnewline \specialrule{0em}{1pt}{1pt}
$d_{L/R}(\tilde g)$ & Representation matrix of $\tilde g$ defined on left/right bond space of MPS. 
\tabularnewline \specialrule{0em}{1pt}{1pt}
\hline 
\hline
\end{tabular}
\end{table}

\begin{table}[H]
\caption{Notations for states and spaces}

\centering{}
\setlength{\tabcolsep}{3mm}
\begin{tabular}{cl}
\hline
\hline 
Symbols & Name meaning \tabularnewline
\hline 
\specialrule{0em}{0pt}{1pt}
$\psi_0$ & Anchor state. 
\tabularnewline \specialrule{0em}{1pt}{1pt}
$\psi_0^{[m]}$ & Anchor state restricted to m adjacent sites(m-cluster). 
\tabularnewline \specialrule{0em}{1pt}{1pt}
$\psi^{[m]}_{\alpha \beta}$ & MPS anchor state restricted to m adjacent sites(m-cluster), with bond index $\alpha$ and $\beta$. 
\tabularnewline \specialrule{0em}{1pt}{1pt}
$\Psi_E$ & Eigen-subspace of $\hat H$ with energy $E$. 
\tabularnewline \specialrule{0em}{1pt}{1pt}
$\Psi_{\tilde G}$ & Subspace generated by anchor state and quasi-symmetry group: $\Psi_{\tilde G}=\mathrm{span}\{\hat{D}(\tilde g)\psi_0|\tilde g\in \tilde G\}$. 
\tabularnewline \specialrule{0em}{1pt}{1pt}
$\Psi_{\tilde G}^{[m]}$ & Subspace generated by $\psi_0^{[m]}$ and quasi-symmetry: $\Psi_{\tilde G}^{[m]} = \mathrm{span}\{\hat{d}^{\otimes m}(\tilde g)\psi_0^{[m]}|\tilde g\in \tilde G\}$. 
\tabularnewline \specialrule{0em}{1pt}{2pt}
\hline 
\hline
\end{tabular}
\end{table}

\section{Appendix B: Product-state as Anchor States}
In this Appendix, we consider one dimension spin-1 system and choose the product state
\begin{eqnarray}
	|\psi_0\rangle = \bigotimes_{i=1}^{N}|+\rangle_i
\end{eqnarray}
as the anchor state. In the following sections, we will discuss 3 possible quasi-symmetry structures and the explicit construction of the scar Hamiltonians.

\subsection{SO(3) Quasi-symmetry}
\label{apndx:PSSO3}
In this section, we consider the SO(3) quasi-symmetry generated by total spin lowing and raising operators:
\begin{eqnarray}
	Q = \sum_{i=1}^{N} S^-_i,\ Q^\dagger = \sum_{i=1}^{N} S^+_i.
\end{eqnarray}
We choose the 2-cluster projector $\hat P_{[j,j+1]}$ projects local 2-cluster states to $S=2$ subspace. The random Hamiltonian
\begin{equation}
\hat{H}=\sum_j(1-\hat{P}_{[j,j+1]})\hat{h}_{[j,j+1]}(1-\hat{P}_{[j,j+1]}),
\end{equation}
is then numerically diagonalized. Level statistics shows the Wigner-Dyson behavior.

\begin{figure}[H]
\begin{centering}
\includegraphics[width=.5\linewidth]{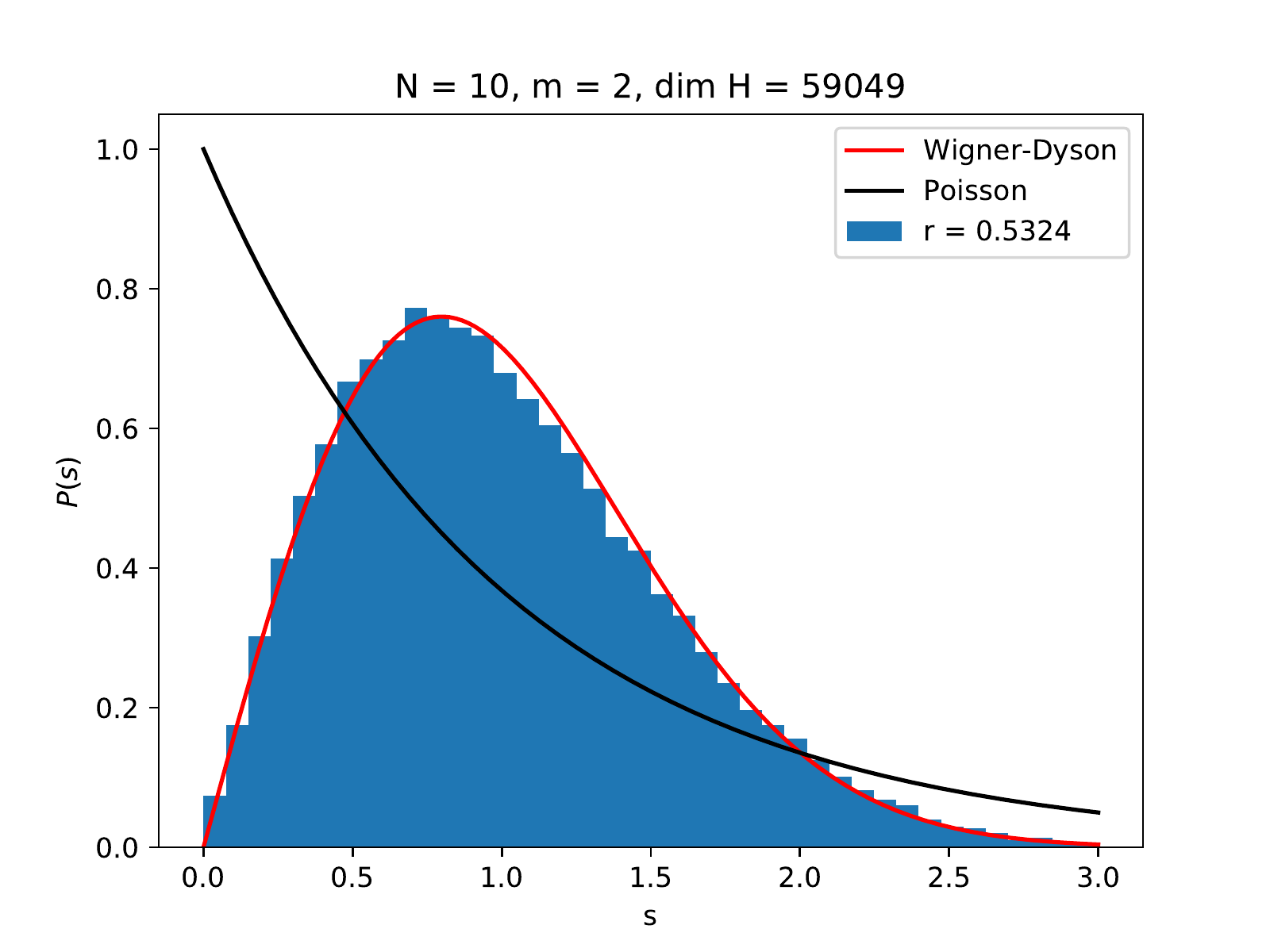}
\par\end{centering}
\caption{Distribution of many-body level spacing $s$ in the middle half of the spectrum of $\hat H$ for $N=10$ with periodic boundary condition. The Hilbert space dimension is 59049. The $r$-statistics is consistent with Wigner-Dyson GOE distribution for chaotic model \cite{PhysRevB.82.174411}.}
\end{figure}

The dimension of the zero-energy subspace (for $3 \le N \le 14$) is numerically found to be
\begin{equation}
\dim \Psi_0 = 2 N + 1.
\end{equation}
On the other hand, the subspace
\begin{equation}
	\Psi_{\tilde{G}} = \mathrm{span}\{\hat{d}^{\otimes 2}(\tilde g) \psi_0^{[2]}|\tilde g \in \tilde G \}\subset \Psi_0
\end{equation}
forms an $S=N$ subspace of the N-chain Hilbert space, which is also of dimension $2N+1$. Therefore, we conclude that up to $N=14$, $\Psi_{\tilde G} = \Psi_0$. Since the subspace $\Psi_0$ holds an irreducible representation of SO(3), without any loss of generality, we can choose a basis for the subspace with respect to total $S_z$ numbers. In this way, the basis become a set of tower states:
\begin{equation}
	\left| \psi_n \right> = \hat Q^n \left| \psi_0 \right>.
\end{equation}

\subsection{SU(2) Quasi-symmetry}
\label{apndx:PSSU2}

Apart from SO(3), there is also an SU(2) subgroup of SU(3), living in the subspace spanned by  $\left\{ \left|+\right\rangle ,\left|-\right\rangle \right\} $, that can be used as a quasi-symmetry. To explicitly define the group, we first define three Pauli operators acting on the $\left\{ \left|+\right\rangle ,\left|-\right\rangle \right\} $ subspace:
\begin{eqnarray}
\hat{\sigma}_{i}^{x} & = & \frac{1}{2}\left[\left(\hat{S}_{i}^{+}\right)^{2}+\left(\hat{S}_{i}^{-}\right)^{2}\right],\\
\hat{\sigma}_{i}^{y} & = & \frac{1}{2i}\left[\left(\hat{S}_{i}^{+}\right)^{2}-\left(\hat{S}_{i}^{-}\right)^{2}\right],\\
\hat{\sigma}_{i}^{z} & = & \hat{S}_{i}^{z}.
\end{eqnarray}
Consider $m=2$, when arbitrary SU(2) elements are acted to local $\left| ++ \right>$ state, the 2-cluster space is then a 3-dimensional space
\begin{equation}
\Psi_{\tilde G}^{[2]}=\mathrm{span}\left\{ \left|--\right\rangle ,\left|+-\right\rangle +\left|-+\right\rangle ,\left|++\right\rangle \right\},
\end{equation}
which forms a 3-dimensional SU(2) representation. The 2-cluster projector is the projection operator to the space $\Psi_{\tilde G}^{[2]}$. The N-chain Hamiltonian
\begin{equation}
\hat{H}= \sum_j\left( 1-\hat{P}_{[j,j+1]} \right) \hat{h}_{[j,j+1]} \left( 1 - \hat{P}_{[j,j+1]} \right)
\end{equation}
is then numerically diagonalized. Level statistics shows Wigner-Dyson behavior.

\begin{figure}[H]
\begin{centering}
\includegraphics[width=.5\linewidth]{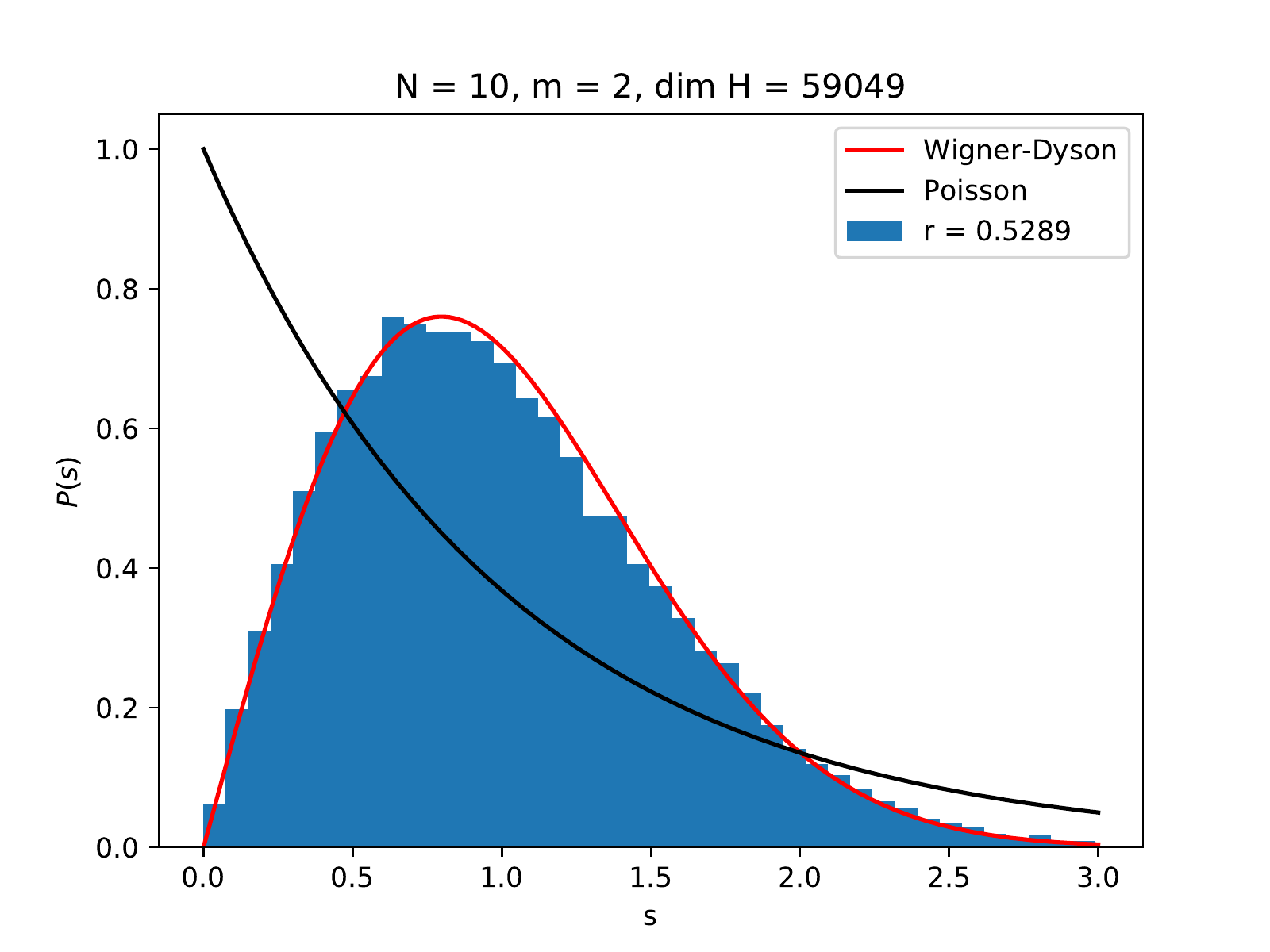}
\par\end{centering}
\caption{Distribution of many-body level spacing $s$ in the middle half of the spectrum of $\hat H$ for $N=10$ with periodic boundary condition. The Hilbert space dimension is 59049. The $r$-statistics is consistent with Wigner-Dyson GOE distribution for chaotic model \cite{PhysRevB.82.174411}.}
\end{figure}

The dimension of zero-energy space (for $3 \le N \le 14$) is numerically found to be
\begin{equation}
	\dim \Psi_0 = N+1
\end{equation}
On the other hand, the subspace $\Psi_{\tilde G} \subset \Psi_0$ forms an ($N+1$)-dimensional irreducible representation of SU(2). In this way, we conclude that up to $N=14$, $\Psi_{\tilde G} = \Psi_0$.

Similarly, the basis for this irreducible representation is also the tower states of the scar model. The explicit form of the tower states is:
\begin{equation}
	\left|\psi_{n}\right\rangle = \hat Q^n \left|\psi_{0}\right\rangle,
\end{equation}
where $\hat Q$ is the lower operator
\begin{equation}
	\hat Q = \sum_{i=1}^N \left( \hat S_i^- \right)^2.
\end{equation}
The type-I scar tower of spin-1 XY model is given by:
\begin{equation}
	\left|\psi_{XY,n}\right\rangle = \left( \hat Q_{XY}^\dagger \right)^n \left|-\cdots-\right\rangle,
\end{equation}
where
\begin{equation}
	\hat Q_{XY} = \sum_{i=1}^N (-1)^i \left( \hat S_i^+ \right)^2.
\end{equation}
These two towers have similar form, and actually they can be related by an onsite unitary transformation
\begin{equation}
\hat{U}=\exp\left(i\frac{\pi}{2}\left(\hat{S}^{z}+1\right)\right)\otimes1\otimes\exp\left(i\frac{\pi}{2}\left(\hat{S}^{z}+1\right)\right)\otimes1\otimes\cdots.
\end{equation}
That is,
\begin{equation}
\left|\psi_{XY,n}\right\rangle =\hat{U}\left|\psi_{N-n}\right\rangle,\ \left|\psi_{n}\right\rangle =\hat{U}\left|\psi_{XY,N-n}\right\rangle.
\end{equation}

\subsection{SU(3) Quasi-symmetry}
\label{apndx:PSSU3}

In Appendix \ref{apndx:PSSO3}-\ref{apndx:PSSU2}, we used SO(3) and SU(2) as the quasi-symmetry groups. In these two simple examples, there is just one ladder operator in the quasi-symmetry group, corresponding to a single set of tower states. A nature question is: can we choose a higher rank Lie group as the quasi-symmetry group? Or more specifically, can we construct a quasi-symmetry model with more than one ladder operators in $\Psi_0$? We give a positive answer to this question by showing a specific model with SU(3) quasi-symmetry.

Consider an SU(3) symmetry that acts fundamentally on the local spins. That is to say, the 3-dimensional local Hilbert space holds a fundamental representation of SU(3), in which 8 generators are represented by 8 Gell-Mann matrices:
$$
\lambda_{1}=\left(\begin{array}{ccc}
0 & 1\\
1 & 0\\
 &  & 0
\end{array}\right),\ \lambda_{2}=\left(\begin{array}{ccc}
0 & -i\\
i & 0\\
 &  & 0
\end{array}\right),\ \lambda_{3}=\left(\begin{array}{ccc}
1\\
 & -1\\
 &  & 0
\end{array}\right),\ \lambda_{4}=\left(\begin{array}{ccc}
 &  & 1\\
 & 0\\
1
\end{array}\right),
$$
\begin{equation}
	\lambda_{5}=\left(\begin{array}{ccc}
	 &  & -i\\
	 & 0\\
	i
	\end{array}\right),\ \lambda_{6}=\left(\begin{array}{ccc}
	0\\
	 & 0 & 1\\
	 & 1 & 0
	\end{array}\right),\ \lambda_{7}=\left(\begin{array}{ccc}
	0\\
	 & 0 & -i\\
	 & i & 0
	\end{array}\right),\ \lambda_{8}=\frac{1}{\sqrt{3}}\left(\begin{array}{ccc}
	1\\
	 & 1\\
	 &  & -2
	\end{array}\right).
\end{equation}
We are free to use different linear combinations of these 8 matrices to form different SU(3) Lie algebras (as a convention, we use capital symbols to represent Lie group, and lower-case symbols to represent Lie algebra). A canonical choice of generators is the Cartan-Weyl basis. In this basis, there are two mutually commuting generators
\begin{equation}
	H_1 = \lambda_8,\ H_2 = \lambda_3,
\end{equation}
which form the Cartan sub-algebra, and 6 ladder operators:
\begin{equation}
	E_1 = \left(\begin{array}{ccc}
 	0 & 1 & \\
	  & 0 & \\
	  &   & 0
	\end{array}\right),\
	F_1 = \left(\begin{array}{ccc}
 	0 &   & \\
	1 & 0 & \\
	  &   & 0
	\end{array}\right),
\end{equation}
\begin{equation}
	E_2 = \left(\begin{array}{ccc}
 	0 &   & \\
	  & 0 & 1\\
	  &   & 0
	\end{array}\right),\
	F_2 = \left(\begin{array}{ccc}
 	0 &   & \\
	  & 0 & \\
	  & 1 & 0
	\end{array}\right),
\end{equation}

\begin{equation}
	E_3 = \left[ E_1,E_2 \right],\
	F_3 = -\left[ F_1,F_2 \right],
\end{equation}
where $E_1,E_2$ are the raising operators corresponding to two simple roots of SU(3). The third ladder can be obtained by commuting $E_1,E_2$ and $F_1,F_2$.

Each irreducible representation of SU(3) is characterized by its highest weight, which is annihilated by both $E_1$ and $E_2$ raising operators. Start from a highest weight, we can get the lower weights by applying $F_1$ and $F_2$ on the obtained weights. This procedure will give us all weight states in this representation. In the current case, the local state $\left| + \right>$ is the highest weight in $(1,0)$ representation. In the language of Lie algebra, we denote such weight state as $\left| (1,0),(1,0) \right>$. Similar as SU(2), the direct product of two highest weights is the highest weight in another representation. This relation can be expressed as:
\begin{equation}
	\left| (M,0),(M,0) \right> \otimes \left| (N,0),(N,0) \right>
	= \left| (M+N,0),(M+N,0) \right>.
\end{equation}
Thus, any m-cluster of the spin all-up state is the highest weight in the $(m,0)$ representation, and the ladder operators in the cluster space are:
\begin{equation}
	\hat Q_1 = \sum_{i=1}^m F_{1,i},\ \hat Q_2 = \sum_{i=1}^m F_{2,i}.
\end{equation}
Now consider a spin-1 chain with spin all-up state as the anchor state, and we choose $m=3$ local cluster. After applying arbitrary SU(3) elements to the anchor state, the 3-cluster space forms a $(3,0)$ representation of SU(3). The $(3,0)$ representation is 10-dimensional. The explicit form of local states are:
\begin{eqnarray}
	\left| (3,0) \right> &=& \left| +++ \right>, \nonumber\\
	\left| (1,1) \right> &=& \left| 0++ \right> +
							  \left| +0+ \right> +
							  \left| ++0 \right>, \nonumber\\
	\left| (\bar 1,2) \right> &=& \left| +00 \right> +
							  \left| 0+0 \right> +
							  \left| 00+ \right>, \nonumber\\
	\left| (\bar 3,3) \right> &=& \left| 000 \right>, \nonumber\\
	\left| (2,\bar 1) \right> &=& \left| -++ \right> +
							  \left| +-+ \right> +
							  \left| ++- \right>, \nonumber\\
	\left| (0,0) \right> &=& \left| +0- \right> +
							  \left| +-0 \right> +
							  \left| 0+- \right> +
							  \left| 0-+ \right> +
							  \left| -+0 \right> +
							  \left| -0+ \right>, \nonumber\\
	\left| (1,\bar 2) \right> &=& \left| +-- \right> +
							  \left| -+- \right> +
							  \left| --+ \right>, \nonumber\\
	\left| (\bar 2,1) \right> &=& \left| -00 \right> +
							  \left| 0-0 \right> +
							  \left| 00- \right>, \nonumber\\
	\left| (\bar 1,\bar 1) \right> &=& \left| 0-- \right> +
							  \left| -0- \right> +
							  \left| --0 \right>, \nonumber\\
	\left| (0,\bar 3) \right> &=& \left| --- \right>,
\end{eqnarray}
where we label the states following the convention of Lie algebra. We then define $\hat P_{[j,j+2]}$ as the 3-cluster projector to $(3,0)$ representation space. The N-chain Hamiltonian
\begin{equation}
	\hat{H}= \sum_j\left( 1-\hat{P}_{[j,j+2]} \right) \hat{h}_{[j,j+2]} \left( 1 - \hat{P}_{[j,j+2]} \right)
\end{equation}
is then numerically diagonalized. Level statistics shows Wigner-Dyson behavior.

\begin{figure}[H]
\begin{centering}
\includegraphics[width=.5\linewidth]{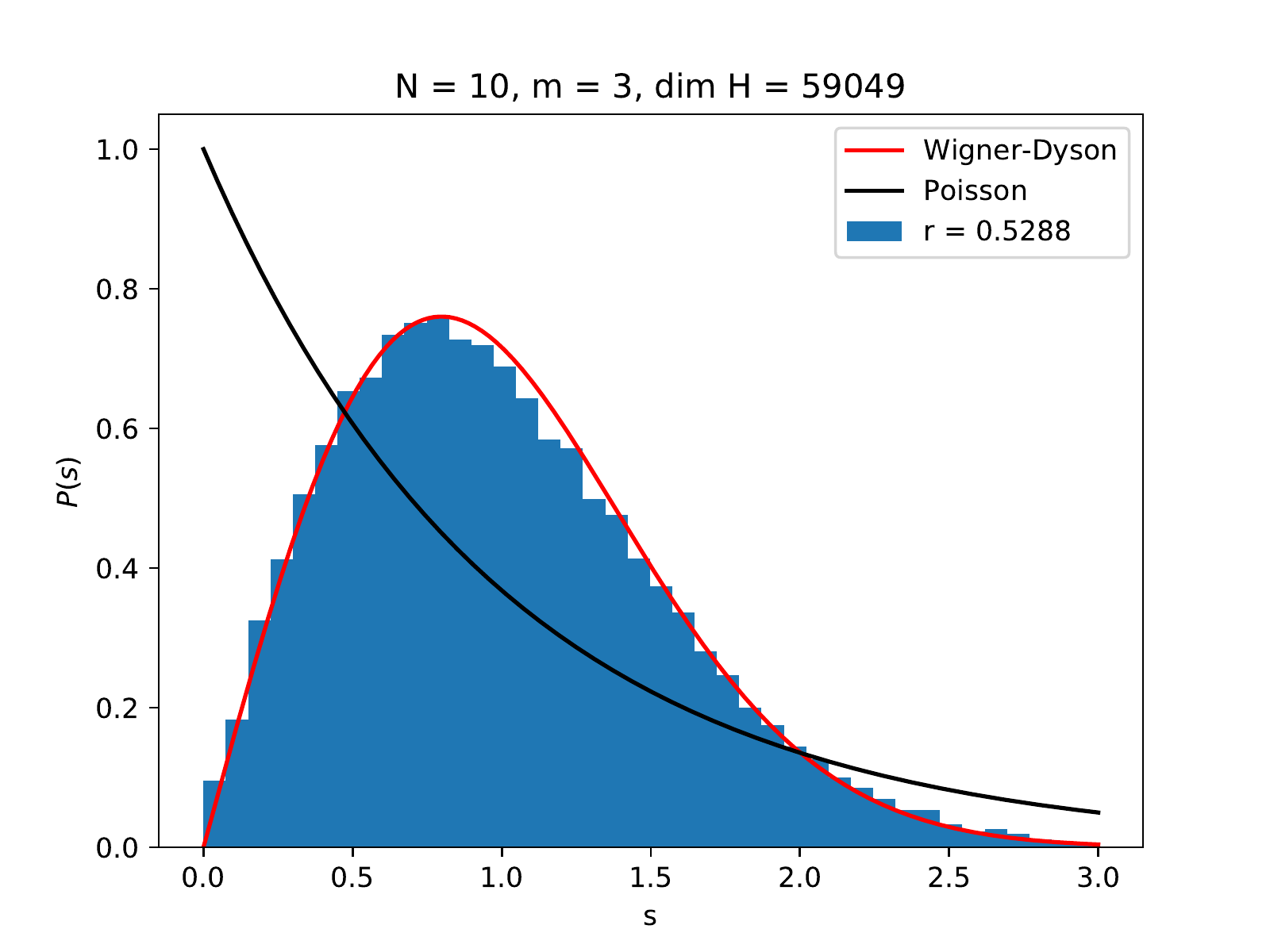}
\par\end{centering}
\caption{Distribution of many-body level spacing $s$ in the middle half of the spectrum of $\hat H$ for $N=10$ with periodic boundary condition. The Hilbert space dimension is 59049. The $r$-statistics is consistent with Wigner-Dyson GOE distribution for chaotic model \cite{PhysRevB.82.174411}.}
\end{figure}

The zero-energy subspace is numerically found (for $4 \le N \le 14$) to have the dimension
\begin{equation}
	\dim \Psi_0 = \frac{(N+1)(N+2)}{2}.
\end{equation}
On the other hand, the group theory analysis shows that $\Psi_{\tilde G}$ is an $(N,0)$ representation space of SU(3), whose dimension is known to be
\begin{equation}
	\dim_{[N,0]} \left( SU(3) \right) = \frac{(N+1)(N+2)}{2}.
\end{equation}
Therefore, we conclude that up to $N=14$, $\Psi_{\tilde G} = \Psi_0$.

Similar as the previous cases, the basis of $(N,0)$ representation can be considered as the scar tower. What different here is that due to the non-commuting ladder operators, the exact structure of the tower states is much more complicated than those in SU(2) quasi-symmetry models. However, in principle, the structure of tower states can all be systematically obtained using the weight diagram technique in Lie algebra.

\section{Appendix C: Matrix-product-state as Anchor State}
\label{apndx:MPS}
In this Appendix, we consider the one dimensional spin-1 system and choose the matrix product state(MPS) with the local matrices
\begin{equation}
A^{[-1]}=\sqrt{\frac{2}{3}}\left(\begin{array}{cc}
1 & 0\\
0 & 0
\end{array}\right),A^{[0]}=\sqrt{\frac{1}{3}}\left(\begin{array}{cc}
0 & 1\\
1 & 0
\end{array}\right),A^{[1]}=\sqrt{\frac{2}{3}}\left(\begin{array}{cc}
0 & 0\\
0 & 1
\end{array}\right)
\label{mps-anchor}
\end{equation}
as the anchor state. In the following sections, we discuss 2 possible quasi-symmetry structures and the explicit construction of the scar Hamiltonians.

\subsection{U(1) Quasi-symmetry}
\label{apndx:MPSU1}

In this section, we discuss the U(1) quasi-symmetry generated by total $S^z$:
\begin{eqnarray}
	\hat S^z = \sum_{i=1}^{N} \hat S_i^z.
\end{eqnarray}
Before proceeding, we first point out that in Ref.\onlinecite{Chattopadhyay2020}, it was found that the type-II scar tower in spin-1 XY model has a projected entangled pairs state (PEPS) structure, in which each spin-1 degree of freedom is split into 2 spin-1/2 degrees of freedom, and each 2 bond-neighboring spin-1/2 degrees of freedom form an entangled pair. In this way, the spin-1 chain can be considered as a ``product state" of entangled pairs, with an onsite projection to the spin-triplet degrees of freedom. The MPS considered here has the similar picture, and can also be brought to a PEPS form:
\begin{equation}
\left|\psi_{0}\right\rangle =\begin{cases}
	\hat P^{(1)}\left(\otimes_{i=1}^{N-1}\left|EP\right\rangle _{2i,2i+1}\right) & OBC\\
	\hat P^{(1)}\left(\otimes_{i=1}^{N-1}\left|EP\right\rangle _{2i,2i+1}\otimes\left|EP\right\rangle _{2N,1}\right) & PBC
	\end{cases},
\end{equation}
where $\hat P^{(1)}$ is the projection to spin-1 degrees of freedom, and the entangled pairs are
\begin{equation}
\left|EP\right\rangle _{2i,2i+1}=\left|\uparrow\uparrow\right\rangle +\left|\downarrow\downarrow\right\rangle =\left(1+s_{i}^{+}s_{i+1}^{+}\right)\left|\downarrow\downarrow\right\rangle .
\end{equation}
Here, the N-site spin-1 chain is split to a 2N-site spin-1/2 chain. We relabel the site indices accordingly, and we use lower-case symbols to represent the spin operators acting on spin-1/2 degrees of freedom.

Since the quasi-symmetry we choose is the abelian group U(1), there is no obvious ladder operators in the zero-energy space. However, we can still group the states according to their total $S_z$ number. It is easier to work on the PEPS picture, and the projector $\hat P^{(1)}$ will not affect the result since it commute with spin operators.

Consider $m=3$ case, the anchor MPS restricted to local 3-cluster is:
\begin{eqnarray}
	\left| \psi_0^{[3]} \right> &=& \hat P^{(1)} \left( \left|s_l\right> \otimes \left|EP\right> \otimes \left|EP\right> \otimes \left|s_r\right> \right) \nonumber \\
	&=& \hat P^{(1)} \left| s_l \uparrow \uparrow \uparrow \uparrow s_r \right> + \hat P^{(1)} \left| s_l \downarrow \downarrow \downarrow \downarrow s_r \right>
	 + \hat P^{(1)} \left( \left| s_l \downarrow \downarrow \uparrow \uparrow s_r \right> + \left| s_l \uparrow \uparrow \downarrow \downarrow s_r \right> \right),
\end{eqnarray}
where $s_l$ and $s_r$ are 2 dangling $S=1/2$ spins and are totally free. In order to have the U(1) quasi-symmetry, the 3-cluster projector we design should preserve each total $S_z$ conserving component of the anchor state. Therefore, we let the 3-cluster space to be:
\begin{equation}
	\Psi_{\tilde G}^{[3]} = \mathrm{span} \{\hat P^{(1)} \left| s_l \uparrow \uparrow \uparrow \uparrow s_r \right>,
					\hat P^{(1)} \left| s_l \downarrow \downarrow \downarrow \downarrow s_r \right>,
	 				\hat P^{(1)} \left( \left| s_l \downarrow \downarrow \uparrow \uparrow s_r \right> +
	 				\left| s_l \uparrow \uparrow \downarrow \downarrow s_r \right> \right) | s_l, s_r=\uparrow, \downarrow \},
\end{equation}
which is a 12-dimensional space, whose explicit basis has been shown in the main text. The N-chain Hamiltonian
\begin{equation}
	\hat{H}= \sum_j\left( 1-\hat{P}_{[j,j+2]} \right) \hat{h}_{[j,j+2]} \left( 1 - \hat{P}_{[j,j+2]} \right)
\end{equation}
is then numerically diagonalized. Level statistics shows Wigner-Dyson behavior. 

\begin{figure}[H]
\begin{centering}
\includegraphics[width=.5\linewidth]{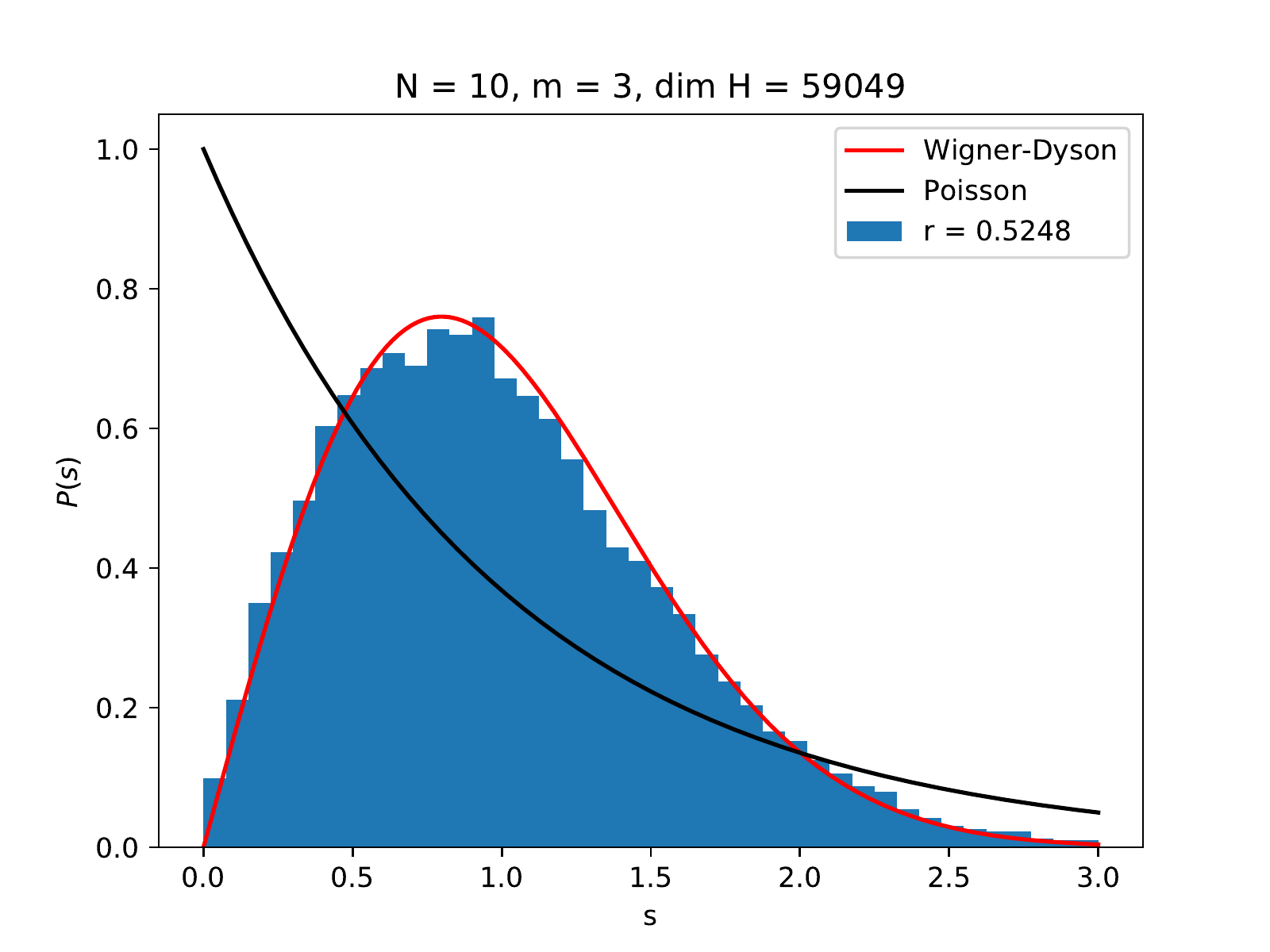}
\par\end{centering}
\caption{Distribution of many-body level spacing $s$ in the middle half of the spectrum of $\hat H$ for $N=10$ with periodic boundary condition. The Hilbert space dimension is 59049. The $r$-statistics is consistent with Wigner-Dyson GOE distribution for chaotic model \cite{PhysRevB.82.174411}.}
\end{figure}

The zero-energy subspace with periodic boundary condition is numerically checked (for $6 \le N \le 14$) to have dimension
\begin{equation}
	\dim \Psi_0 = N+1.
\end{equation}
While for the case with open boundary condition, it is numerically found (for $5 \le N \le 14$) that
\begin{equation}
	\dim \Psi_0 = 4L.
\end{equation}
On the other hand, our previous analysis shows that the quasi-symmetry space is spanned by a set of tower states with a ``hidden ladder operator" acting on the entangled pairs:
\begin{equation}
\hat{Q}^{\dagger}=\begin{cases}
\sum_{i=1}^{N-1}s_{2i}^{+}s_{2i+1}^{+} & OBC\\
\sum_{i=1}^{N-1}s_{2i}^{+}s_{2i+1}^{+}+s_{2N}^{+}s_{1}^{+} & PBC
\end{cases}
\end{equation}
The lowest state in the tower is:
\begin{equation}
\left|\phi_{0}\right\rangle =\begin{cases}
	\hat P^{(1)}\left|s_{l}\downarrow\cdots\downarrow s_{r}\right\rangle  & OBC\\
	\hat P^{(1)}\left|\downarrow\cdots\downarrow\right\rangle  & PBC
\end{cases}.
\end{equation}
And by acting ladder operator on the entangled pairs, we can get all other tower states:
\begin{equation}
\left|\phi_{n}\right\rangle =\hat P^{(1)} \left(\hat{Q}^{\dagger}\right)^{n}\left|\phi_0\right\rangle .
\end{equation}
The tower state can be recombined to be the initial anchor state:
\begin{equation}
\left|\psi_{0}\right\rangle = \sum_{n=0}^{N} C_n \left|\phi_{n}\right\rangle.
\end{equation}
In this way, we show that the dimension of $\Psi_{\tilde G}$ is
\begin{equation}
\dim \Psi_{\tilde G}=\begin{cases}
N+1 & PBC\\
4N & OBC
\end{cases}.
\end{equation}
Therefore, We conclude that up to $N=14$, $\Psi_{\tilde G} = \Psi_0$.

Finally, we note that the scar tower in this model can be mapped to type-II scar tower in spin-1-XY model by an onsite unitary operator $\hat{U}$:
\begin{equation}
\left|\psi_{XY,n}\right\rangle =\hat{U}\left|\phi_{n}\right\rangle ,
\end{equation}
where $\hat{U}$ is a chain operator with period of 4 sites:
\begin{equation}
\hat{U}=\left[\exp\left(i\pi \hat S^{z}\right)\otimes\exp\left(i\pi \hat S^{z}\right)\otimes1\otimes1\right]\otimes\cdots.
\end{equation}

\subsection{SO(3) Quasi-symmetry}
\label{apndx:MPSSO3}

In this section, we show that a non-abelian quasi-symmetry can be imposed on the MPS anchor state. The quasi-symmetry we choose is the SO(3) spin rotation. It is numerically found that if we require $\Psi_{\tilde G} = \Psi_0$, we should choose $m \ge 4$. In this way, the projected space is then
\begin{equation}
	\Psi_{\tilde G}^{[4]} = \mathrm{span}\left\{\hat{d}^{\otimes 4}(\tilde g) \psi_0^{[4]}|\tilde g \in \mathrm{SO(3)} \right\}.
\end{equation}
The subspace $\Psi_{\tilde G}^{[4]}$ is numerically obtained, whose dimension is
\begin{equation}
	\dim\Psi_{\tilde G}^{[4]}=40.
\end{equation}
We define $\hat P_{[j,j+3]}$ as the projection to the local 4-cluster space. The N-chain Hamiltonian
\begin{equation}
	\hat{H}= \sum_j\left( 1-\hat{P}_{[j,j+3]} \right) \hat{h}_{[j,j+3]} \left( 1 - \hat{P}_{[j,j+3]} \right)
\end{equation}
is then numerically diagonalized. Level statistics shows Wigner-Dyson behavior.

\begin{figure}[H]
\begin{centering}
\includegraphics[width=.5\linewidth]{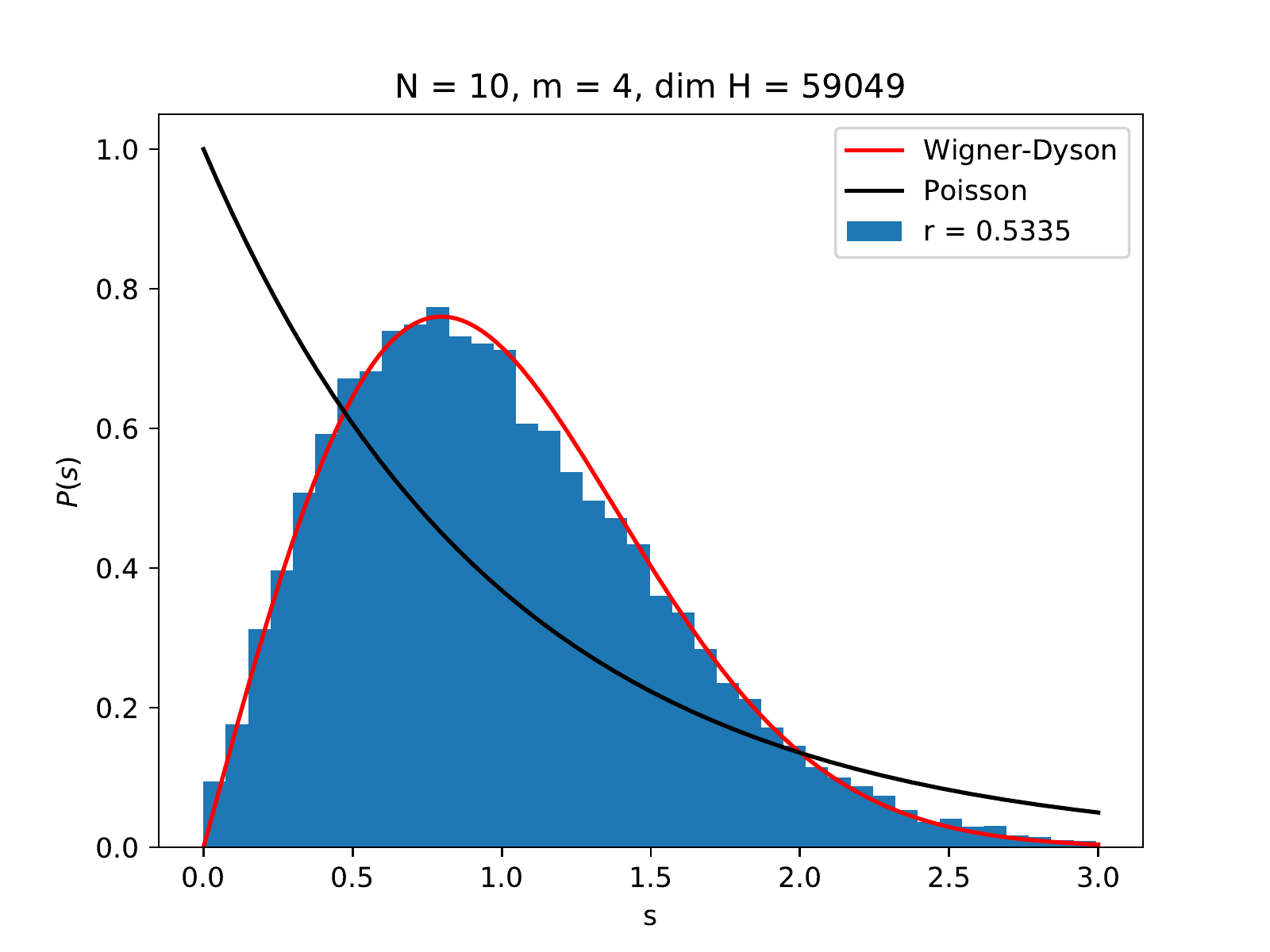}
\par\end{centering}
\caption{Distribution of many-body level spacing $s$ in the middle half of the spectrum of $\hat H$ for $N=10$ with periodic boundary condition. The Hilbert space dimension is 59049. The $r$-statistics is consistent with Wigner-Dyson GOE distribution for chaotic model \cite{PhysRevB.82.174411}.}
\end{figure}

The zero-energy space in open boundary condition is numerically found (for $5 \le N \le 14$) to be
\begin{equation}
	\dim \Psi_0 = 2 N (N+1).
\end{equation}
We can also group the states in this subspace according to their total S numbers. This is done by numerically diagonalize the $\hat S^2$ operator under the basis of $\Psi_0$. The result is listed below:

\begin{table}[H]
\caption{Classification of zero-energy states with open boundary condition by the total spin number. The numbers in the table represent the multiplicity of each irreducible SO(3) representation in the zero-energy subspace of the Hamiltonian.}

\centering{}
\setlength{\tabcolsep}{3mm}
\begin{tabular}{cccccccccccccc}
\hline
\hline 
$S$ & 0 & 1 & 2 & 3 & 4 & 5 & 6 & 7 & 8 & 9 & 10 & 11 & 12\tabularnewline
\hline 
$N=4$ & 1 & 2 & 2 & 2 & 1 &  &  &  &  &  &  &  & \tabularnewline
$N=5$ & 1 & 2 & 2 & 2 & 2 & 1 &  &  &  &  &  &  & \tabularnewline
$N=6$ & 1 & 2 & 2 & 2 & 2 & 2 & 1 &  &  &  &  &  & \tabularnewline
$N=7$ & 1 & 2 & 2 & 2 & 2 & 2 & 2 & 1 &  &  &  &  & \tabularnewline
$N=8$ & 1 & 2 & 2 & 2 & 2 & 2 & 2 & 2 & 1 &  &  &  & \tabularnewline
$N=9$ & 1 & 2 & 2 & 2 & 2 & 2 & 2 & 2 & 2 & 1 &  &  & \tabularnewline
$N=10$ & 1 & 2 & 2 & 2 & 2 & 2 & 2 & 2 & 2 & 2 & 1 &  & \tabularnewline
$N=11$ & 1 & 2 & 2 & 2 & 2 & 2 & 2 & 2 & 2 & 2 & 2 & 1 & \tabularnewline
$N=12$ & 1 & 2 & 2 & 2 & 2 & 2 & 2 & 2 & 2 & 2 & 2 & 2 & 1\tabularnewline
\hline 
\hline
\end{tabular}
\end{table}

While for the periodic boundary condition, it is numerically found (for $5 \le N \le 14$) that
\begin{equation}
	\dim \Psi_0 = \frac{(N+1)(N+2)}{2}.
\end{equation}
Similarly, we can group the states according to their total S numbers. The numerical result is listed below:

\begin{table}[H]
\caption{Classification of zero-energy states with periodic boundary condition by the total spin number. The numbers in the table represent the multiplicity of each irreducible SO(3) representation in the zero-energy subspace of the Hamiltonian.}

\centering{}
\setlength{\tabcolsep}{3mm}
\begin{tabular}{cccccccccccccc}
\hline
\hline 
$S$ & 0 & 1 & 2 & 3 & 4 & 5 & 6 & 7 & 8 & 9 & 10 & 11 & 12\tabularnewline
\hline 
$N=4$ & 1 &  & 1 &  & 1 &  &  &  &  &  &  &  & \tabularnewline
$N=5$ &  & 1 &  & 1 &  & 1 &  &  &  &  &  &  & \tabularnewline
$N=6$ & 1 &  & 1 &  & 1 &  & 1 &  &  &  &  &  & \tabularnewline
$N=7$ &  & 1 &  & 1 &  & 1 &  & 1 &  &  &  &  & \tabularnewline
$N=8$ & 1 &  & 1 &  & 1 &  & 1 &  & 1 &  &  &  & \tabularnewline
$N=9$ &  & 1 &  & 1 &  & 1 &  & 1 &  & 1 &  &  & \tabularnewline
$N=10$ & 1 &  & 1 &  & 1 &  & 1 &  & 1 &  & 1 &  & \tabularnewline
$N=11$ &  & 1 &  & 1 &  & 1 &  & 1 &  & 1 &  & 1 & \tabularnewline
$N=12$ & 1 &  & 1 &  & 1 &  & 1 &  & 1 &  & 1 &  & 1\tabularnewline
\hline
\hline 
\end{tabular}
\end{table}

And we also numerically checked that up to $N=14$, $\Psi_{\tilde G}=\Psi_{0}$.

\section{Appendix D: General Discussions}

\subsection{Discrete Quasi-symmetry}
\label{apndx:disqs}

In principle, the construction scheme introduced in the main text also works for cases where the quasi-symmetry group is discrete.
We assume that the group is finite.
Following exactly the same steps in the main text, one can obtain a non-integrable $\hat{H}$ that has discrete $\tilde{G}$ for its quasi-symmetry group with respect to the zero-energy subspace.
However, if the anchor state $\psi_0$ is not carefully chosen, there is sometimes a larger Lie group $\tilde{G}'\supset\tilde{G}$, under the action of which the zero-energy subspace is invariant.
This would imply that the real quasi-symmetry group is $\tilde{G}'$ and this takes us back to the Lie quasi-symmetry groups again.
For example, assume $s=1$, and suppose $\psi_0=|\mathbf{n}\rangle\otimes\dots\otimes|\mathbf{n}\rangle$ where $\mathbf{n}$, where $|\mathbf{n}\rangle$ is a fully polarized spin along some generic direction $\mathbf{n}$ in 3-space, and choose $\tilde{G}$ to be the cubic group $O$ (all proper rotations of a cube).
There are 24 elements in $O$, so acting them on a $2$-cluster of $\psi_0$ results in 24 states, if $\mathbf{n}$ is a generic direction.
Here the ``generic direction'' is defined as a direction such that there is not any subgroup $1\subset\tilde{H}\subset\tilde{G}$, under which this direction is invariant.
Out of the 24 states that span $\Psi_{\tilde{G}}^{[2]}$, we show by explicit calculation that there are always five independent states, exactly forming the $S=2$ SO(3) subspace of two spins.
Therefore, the $2$-cluster projector onto $\Psi_{\tilde{G}}^{[2]}$ is in fact SO(3) invariant, so is the total $\hat{H}$ constructed from these projectors.

We do not know, in general, how to avoid this ``emergent Lie groups'' in our construction of discrete quasi-symmetry groups.
But we do have examples when the resultant group remains a discrete one, by choosing an anchor state polarized in a high-symmetry direction.
Again consider $s=1$ and $m=2$.
Then consider $\psi_0=|z\dots{z}\rangle$, where $|z\rangle$ is the eigenstate of $\hat{S}_z$ with zero eigenvalue.
Acting all elements of $O$ on a $2$-cluster $|zz\rangle$ results in a subspace spanned by $|xx\rangle, |yy\rangle, |zz\rangle$.
Notice that these three states do \emph{not} form the $S=1$ subspace of SO(3).
The $2$-cluster projector then is
\begin{equation}
\hat{P}=|xx\rangle\langle{xx}|+|yy\rangle\langle{yy}|+|zz\rangle\langle{zz}|.
\end{equation}
The $N$-chain Hamiltonian
\begin{equation}
\hat{H}=\sum_j(1-\hat{P}_{[j,j+1]})\hat{h}_{[j,j+1]}(1-\hat{P}_{[j,j+1]})
\end{equation}
is numerically diagonalized.
Level statistics shows the Wigner-Dyson behavior.
\begin{figure}[H]
\begin{centering}
\includegraphics[width=.5\linewidth]{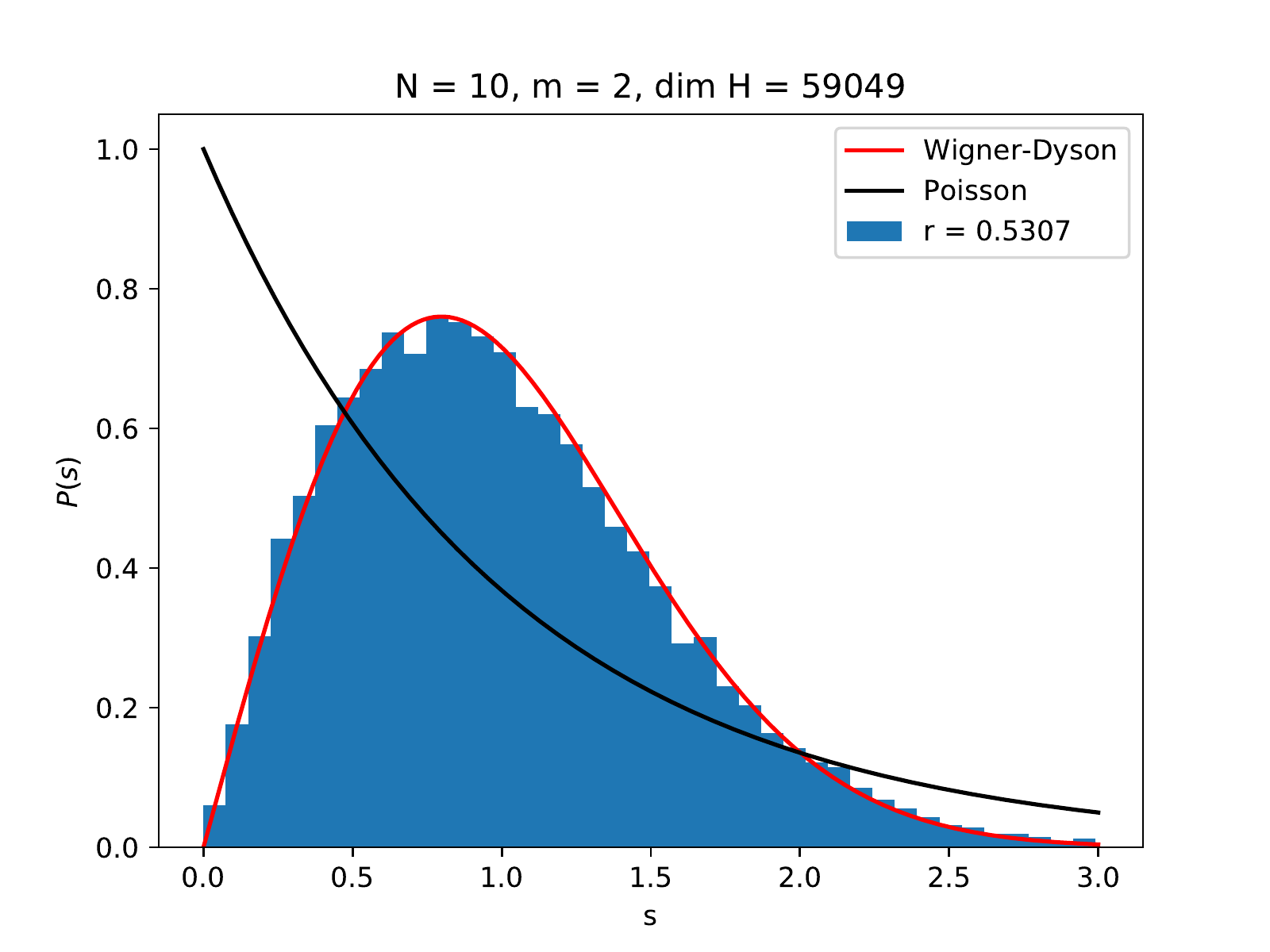}
\par\end{centering}
\caption{Distribution of many-body level spacing $s$ in the middle half of the spectrum of $\hat H$ for $N=10$ with periodic boundary condition. The Hilbert space dimension is 59049. The $r$-statistics is consistent with Wigner-Dyson GOE distribution for chaotic model \cite{PhysRevB.82.174411}.}
\end{figure}
Here dimension of the zero-energy subspace does not scale with $N$ as in all the Lie-group cases, but remains $\dim(\Psi_0)=3$ for $N\le14$.
The three zero energy states are just $|x\dots{x}\rangle, |y\dots{y}\rangle$ and $\psi_0=|z\dots{z}\rangle$.

For $\tilde{G}$ being discrete, there is not an obvious way of finding $\hat{H}_1$, such that $\hat{H}_{scar}=\hat{H}+\hat{H}_1$, shows exact periodicity in the time evolution of $\psi\in\Psi_0$.
In this case, we consider a discrete time series $t_{1,2,\dots}$.
At $t=t_i$, an onsite unitary operator $\hat{D}(\tilde{g})$, where $\tilde{g}\in\tilde{G}$, is applied to the state, and at $t\neq{t}_i$, the state evolves under $\hat{H}$.
This pulse evolution can be realized by the following Hamiltonian
\begin{equation}
\hat{H}_{scar}=\hat{H}+\sum_i\delta(t-t_i)\log(\hat{D}(\tilde{g})).
\end{equation}
Since $\tilde{G}$ is finite, so each element $\tilde{g}$ has a definite order $\tilde{g}^{n_{\tilde{g}}}=1$, after $n_{\tilde{g}}$ pulses, the state vector goes back to itself, or symbolically
\begin{equation}
\psi(t)\propto\psi(t'),
\end{equation}
if $t\in(t_j,t_{j+1})$ and $t'\in(t_{j+n_{\tilde{g}}},t_{j+n_{\tilde{g}}+1})$.

\subsection{Inhomogeneous Anchor States and Group Actions}
\label{apndx:inhomo}

In the main text, we have used homogeneous anchor states, and the actions of quasi-symmetry group are also homogeneous.
For a product state as the anchor state, we have assumed that $\psi_0=\phi\otimes\dots\otimes\phi$ where $\phi$ is a single-spin state.
For a matrix-product state as the anchor state, we have assumed that $\psi_0=\sum_{s_1,\dots,s_N}Tr[A^{s_1}\dots{A}^{s_N}]|s_1,\dots,s_N\rangle$.
The group actions are assumed to take the form $\hat{D}(\tilde{g})=\hat{d}^{\otimes{N}}(\tilde{g})$.

These assumptions can be relaxed to some degree.
The group actions can be extended to the form where the representation $\hat{d}_i(\tilde G)$ is site-dependent, i.e., in the definition
\begin{equation}
\hat{D}(\tilde{g})=\hat{d}_1(\tilde{g})\otimes\dots\otimes \hat{d}_N(\tilde{g}),
\end{equation}
we let $\hat{d}_i(\tilde{g})$'s be $N$ different representations of $\tilde{G}$.
The form of the product-anchor state can also be extended to
\begin{equation}
\psi_0=\phi_1\otimes\dots\otimes\phi_N,
\end{equation}
such that there is $\tilde g_j \in \tilde G$
\begin{equation}
\phi_j=\hat{d}_j(\tilde g_j)\phi_1.
\end{equation}
The matrix-product-anchor state can be extended similarly, where
\begin{equation}
A^{s}\rightarrow{A}^{s}_{j}=[d_j(\tilde g_j)]_{ss'}A^{s'}_j,
\end{equation}
where $d_j(\tilde G)$ is the site-dependent linear representation of $\tilde G$.

After these extension, the form of $m$-cluster projector $\hat{P}_{[j,j+m-1]}$ now depends on $j$, obtained after the following process.
For the chain $[j,j+m-1]$, define
\begin{eqnarray}
\psi_0^{[j,j+m-1]}&\equiv&\phi_j\otimes\dots\otimes\phi_{j+m-1},\\
\left[\otimes_{i=j}^{j+m-1}\hat{d}(\tilde{g})\right]\psi_0^{[j,j+m-1]}&\equiv&\hat{d}_j(\tilde{g})\phi_j\otimes\dots\otimes\hat{d}_{j+m-1}(\tilde{g})\phi_{j+m-1}.
\end{eqnarray}
$\Psi_{\tilde G}^{[j,j+m-1]}$ is then defined as the Hilbert subspace spanned by all $[\otimes_{i=j}^{j+m-1}\hat{d}(\tilde{g})]\psi_0^{[j,j+m-1]}$, and $\hat{P}_{[j,j+m-1]}$ is the $m$-cluster projector onto this subspace
With the projectors defined, the full Hamiltonian is given by the same equation as in the main text
\begin{equation}
\hat{H}=\sum_j(1-\hat{P}_{[j,j+m-1]})\hat{h}_{[j,j+m-1]}(1-\hat{P}_{[j,j+m-1]}).
\end{equation}

\subsection{Non-abelian Quasi-symmetry and Ladder Operators}
For a Hamiltonian $\hat{H}$ having quasi-symmetry $\tilde{G}$ with respect to $\Psi_0$, we have
\begin{equation}
[\hat{H},\hat{D}(\tilde{g})]\Psi_0=0,\ \forall \tilde g \in \tilde G.
\end{equation}
If $\tilde{G}$ is non-abelian, there is at least one SO(3) subgroup $\tilde{G}\supset\tilde{W}\cong\mathrm{SO(3)}$.
Consider the generators of $\tilde{W}$, $\hat{L}_{x,y,z}$, we have
\begin{equation}\label{eq:a1}
[\hat{H},\hat{L}_+]\Psi_0=[\hat{H},\hat{L}_-]\Psi_0=0,
\end{equation}
and
\begin{equation}\label{eq:a2}
[\hat{L}_z,\hat{L}_\pm]= \pm \hat{L}_\pm,
\end{equation}
where $\hat{L}_{\pm}\equiv\hat{L}_x\pm{i}\hat{L}_y$.
Now, construct a scar Hamiltonian following the main text, $\hat{H}_{scar}=\hat{H}+c\hat{L}_z$. Eq.(\ref{eq:a1},\ref{eq:a2}) give us
\begin{equation}
[\hat{H}_{scar},\hat{L}_{\pm}]\Psi_0=\pm{c}\hat{L}_\pm\Psi_0.
\end{equation}
Therefore, any non-abelian quasi-symmetry group implies the existence of what we call a two-way-ladder operator.
A two-way-ladder operator $\hat Q_2$, for a given Hamiltonian $\hat{H}$ and a subspace $\Psi$, is defined as an operator such that
\begin{eqnarray}
[\hat{H},\hat Q_2^\dag]\Psi &=& c \hat Q_2^\dag \Psi,\\
\nonumber
[\hat{H},\hat Q_2]\Psi&=&-c\hat Q_2\Psi.
\end{eqnarray}

However, not all scar models discussed in literature has a two-way-ladder operator.
(For reference, the type-I scar tower in spin-1-XY model \cite{Schecter2019} and the $\eta$-pairing scar tower in Hubbard model \cite{Moudgalya2020a,Mark2020a} have.)
Some scar towers only have a one-way ladder $\hat{Q}_1$ such that
\begin{equation}
[\hat{H},\hat Q_1^\dag]\Psi=c\hat Q_1^\dag\Psi.
\end{equation}
Naturally, these scar towers only have U(1)-quasi-symmetry, simply because any non-abelian quasi-symmetry would imply a two-way-ladder operator.
The statement does not go reversely.
A U(1)-quasi-symmetry does not imply the existence of a one-way-ladder operator: the type-II scar tower in spin-1-XY model \cite{Schecter2019,Chattopadhyay2020} is such an example.

Below we work out an explicit example, where we use the construction scheme in the main text to generate a scar tower that is very close to the Affleck-Kennedy-Lieb-Tasaki(AKLT)-tower as appears in Ref.\onlinecite{Moudgalya2018}.

The AKLT state has the following MPS representation:
\begin{equation}
	\left| AKLT \right> = \sum_{\left\{ i_{n}\right\} }
	Tr \left(A^{[i_1]}A^{[i_2]}\cdots A^{[i_N]}\right)
	\left|i_1,i_2\cdots,i_N\right\rangle,
\end{equation}
\begin{equation}
	A^{\left[+\right]}=\sqrt{\frac{2}{3}}\left(
	\begin{array}{cc}
	0 & 1\\
	0 & 0
	\end{array}\right),
	A^{\left[0\right]}=\sqrt{\frac{1}{3}}\left(
	\begin{array}{cc}
	-1 & 0\\
	0 & 1
	\end{array}\right),
	A^{\left[-\right]}=\sqrt{\frac{2}{3}}\left(
	\begin{array}{cc}
	0 & 0\\
	-1 & 0
	\end{array}\right).
\end{equation}
In Ref.\onlinecite{Moudgalya2018}, a set of exact scar tower was found for AKLT model:
\begin{equation}
	\left| \phi_n \right> = \left(\hat Q_1^\dagger \right)^n \left| AKLT \right>,
\end{equation}
where
\begin{equation}
	\hat Q_1^\dagger = \sum_j (-1)^j (\hat S^+_j )^2, \label{eq:akltladder}
\end{equation}
The scar tower starts from AKLT state and ends with ferromagnetic state (while for $N=4L+2$, the ferromagnetic state cannot be reached \cite{Mark2020})

In Ref.\onlinecite{Mark2020}, a superposition of the tower states was found which is a low entangled initial state with revival dynamics
\begin{equation}
	\left| \psi_0 \right> = \exp( \alpha \hat Q_1^\dagger) \left| AKLT \right>.
\end{equation}
While in Ref.\onlinecite{Mark2020}, the operator $\exp( \alpha \hat Q_1^\dagger)$ was represented by a $4\times4$ matrix-product-operator(MPO). When applying the MPO to AKLT MPS, the MPO$\times$MPS gives an $8 \times 8$ MPS. Here we take a different view by noticing that Eq.(\ref{eq:akltladder}) is the sum of onsite operator. The exponential of the summation is then a direct product of onsite operators:
\begin{equation}
	\exp(\alpha \hat Q_1^\dagger) = \otimes_{j=1}^N \left[1+(-1)^j\alpha (\hat S^+_j)^2 \right].
\end{equation}
This onsite operator, when acting on MPS, will not change the MPS's bond dimension. The specific form of $\left| \psi_0 \right>$ is
\begin{equation}
	\left| \psi_0 \right> = \sum_{\left\{ i_{n}\right\} }
	Tr \left(B^{[i_1]}C^{[i_2]}\cdots B^{[i_{N-1}]}C^{[i_N]}\right)
	\left|i_1,i_2\cdots,i_{N-1},i_N\right\rangle,
\end{equation}
where
\begin{equation}
	B^{\left[+\right]}=\sqrt{\frac{2}{3}}\left(
	\begin{array}{cc}
	0 & 1\\
	\alpha & 0
	\end{array}\right),
	B^{\left[0\right]}=\sqrt{\frac{1}{3}}\left(
	\begin{array}{cc}
	-1 & 0\\
	0 & 1
	\end{array}\right),
	B^{\left[-\right]}=\sqrt{\frac{2}{3}}\left(
	\begin{array}{cc}
	0 & 0\\
	-1 & 0
	\end{array}\right),
\end{equation}
\begin{equation}
	C^{\left[+\right]}=\sqrt{\frac{2}{3}}\left(
	\begin{array}{cc}
	0 & 1\\
	-\alpha & 0
	\end{array}\right),
	C^{\left[0\right]}=\sqrt{\frac{1}{3}}\left(
	\begin{array}{cc}
	-1 & 0\\
	0 & 1
	\end{array}\right),
	C^{\left[-\right]}=\sqrt{\frac{2}{3}}\left(
	\begin{array}{cc}
	0 & 0\\
	-1 & 0
	\end{array}\right).
\end{equation}
The value of $\alpha$ can be arbitrary (as long as it is non-zero). Here we set $\alpha = 1$ for concreteness. We then choose $\left| \psi_0 \right>$ as the anchor state and spin rotation along z-axis as the quasi-symmetry. And choose ($m=3$)-cluster for the local projector. The projected space
\begin{equation}
	\Psi_{\tilde G}^{[3]} = \left\{\hat{d}^{\otimes 3}(\tilde g) \psi_0^{[3]}|\tilde g \in \mathrm{U(1)} \right\}
\end{equation}
is found to be 8-dimensional, spanned by the following 8 basis states:
\begin{eqnarray}
	\left| \psi^{[3]}_1 \right> &=& \left| +++ \right>, \nonumber\\
	\left| \psi^{[3]}_2 \right> &=& \left| +0+ \right>, \nonumber\\
	\left| \psi^{[3]}_3 \right> &=& \left| ++0 \right> -
							  \left| 0++ \right>, \nonumber\\
	\left| \psi^{[3]}_4 \right> &=& \left| ++- \right> +
							  \left| -++ \right> -
							  \left| +-+ \right> -
							  \left| 0+0 \right>, \nonumber\\
	\left| \psi^{[3]}_5 \right> &=& 5\left| +-+ \right> +
							  3\left| ++- \right> +
							  3\left| -++ \right> +
							  \left| 0+0 \right> -
							  4\left| +00 \right> -
							  4\left| 00+ \right>, \nonumber\\
	\left| \psi^{[3]}_6 \right> &=& \left| +-0 \right> +
							  \left| -0+ \right> +
							  \left| 0+- \right> -
							  \left| -+0 \right> -
							  \left| +0- \right> -
							  \left| 0-+ \right>, \nonumber\\
	\left| \psi^{[3]}_7 \right> &=& \left| +-0 \right> +
							  \left| 0+- \right> +
							  \left| -+0 \right> +
							  \left| 0-+ \right> -
							  \left| -0+ \right> -
							  \left| +0- \right> -
							  \left| 000 \right>, \nonumber\\
	\left| \psi^{[3]}_8 \right> &=& 2\left| -+- \right> +
							  \left| 0-0 \right> -
							  \left| 00- \right> -
							  \left| -00 \right>.
	\end{eqnarray}
We then follow the same procedure. Define $\hat P_{[j,j+2]}$ as the projector to the projected space, and the many-body Hamiltonian as
\begin{equation}
	\hat{H}= \sum_j\left( 1-\hat{P}_{[j,j+2]} \right) \hat{h}_{[j,j+2]} \left( 1 - \hat{P}_{[j,j+2]} \right). \label{eq:akltscarham}
\end{equation}
The $\hat h_{[j,j+1]}$ is chosen to be random Hermitian. The Hamiltonian for $N=10$ with periodic boundary condition is then numerically diagonalized. Level statistics shows Wigner-Dyson behavior.

\begin{figure}[H]
\begin{centering}
\includegraphics[width=.5\linewidth]{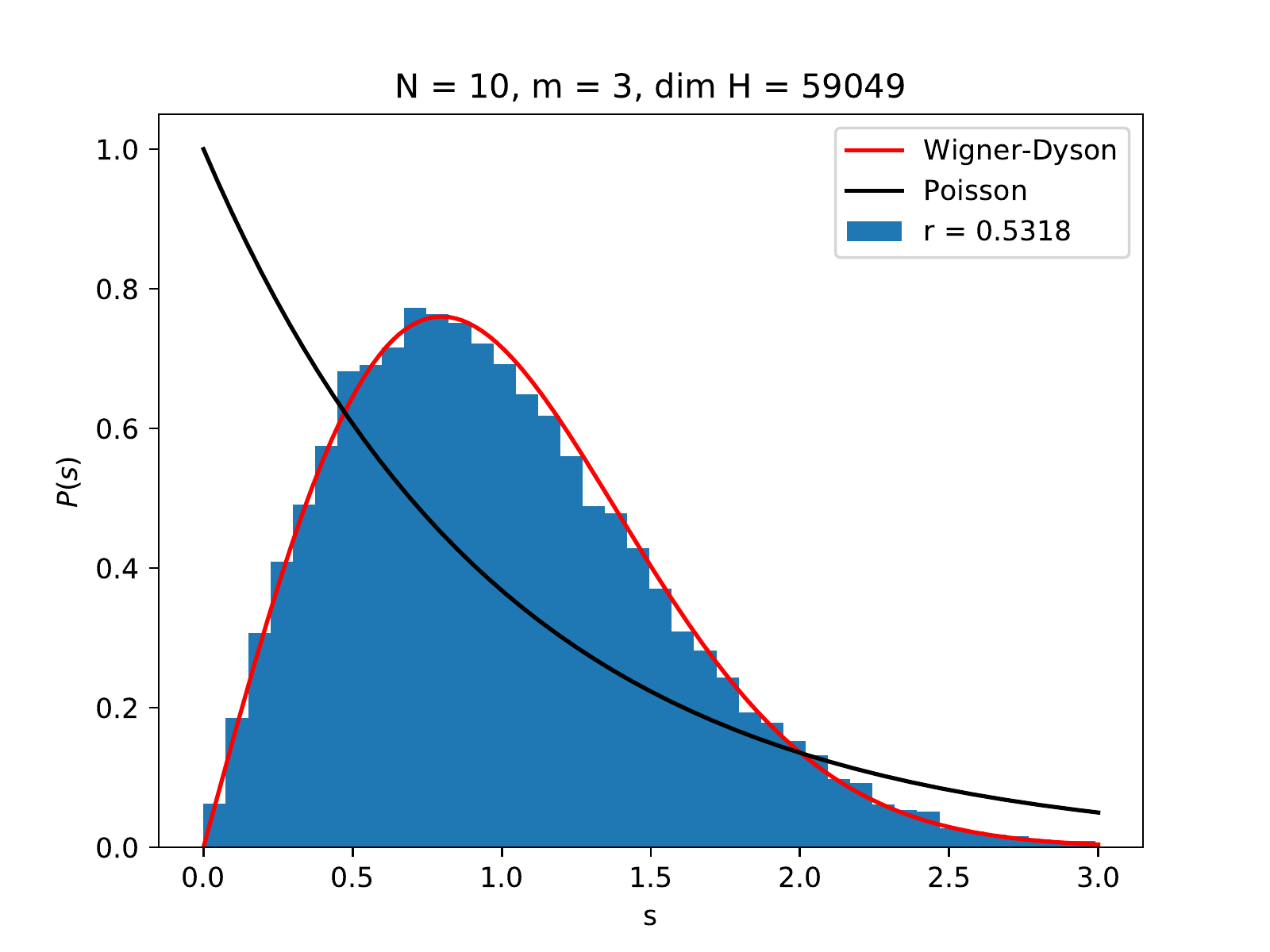}
\par\end{centering}
\caption{Distribution of many-body level spacing $s$ in the middle half of the spectrum of $\hat H$ for $N=10$ with periodic boundary condition. The Hilbert space dimension is 59049. The $r$-statistics is consistent with Wigner-Dyson GOE distribution for chaotic model \cite{PhysRevB.82.174411}.}
\end{figure}

We then numerically find the dimension of zero-energy space of Eq.(\ref{eq:akltscarham}) with periodic boundary condition for ($4 \le N \le 14$), the result is
\begin{equation}
\dim \Psi_0=\begin{cases}
	\frac{N}{2}+3 & N=4L\\
	\frac{N}{2}+1 & N=4L+2
\end{cases}.
\end{equation}
However, the number of state in this scar tower is \cite{Mark2020} (in this case, the number of tower states is exactly the dimension of subspace $\Psi_{\tilde G}$):
\begin{equation}
N_{tower}=
\dim \Psi_{\tilde G}=\begin{cases}
	\frac{N}{2}+1 & N=4L\\
	\frac{N}{2} & N=4L+2
\end{cases}.
\end{equation}
In this case $\Psi_{\tilde G} \subset \Psi_0$. We also numerically checked that this discrepancy cannot be removed by simply enlarging the m-cluster. To see which states are added to the zero-energy sector, we first compute the total $S^z$ number of the zero-energy states. The result is listed below:

\begin{table}[H]
\caption{Classification of zero-energy states with open boundary condition by the total $S^z$ number. The numbers in the table represent the number of states in each total $S^z$ sector.}
\centering{}
\setlength{\tabcolsep}{3mm}
\begin{tabular}{cccccccccccccccc}
\hline
\hline 
$S^z$ & 0 & 1 & 2 & 3 & 4 & 5 & 6 & 7 & 8 & 9 & 10 & 11 & 12 & 13 & 14\tabularnewline
\hline 
$N= 4$ & 1 &   & 1 & 2 & 1 &   &   &   &   &   &   &   &   &  &\tabularnewline
$N= 6$ & 1 &   & 1 &   & 1 &   & 1 &   &   &   &   &   &   &  & \tabularnewline
$N= 8$ & 1 &   & 1 &   & 1 &   & 1 & 2 & 1 &   &   &   &   &  & \tabularnewline
$N=10$ & 1 &   & 1 &   & 1 &   & 1 &   & 1 &   & 1 &   &   &  & \tabularnewline
$N=12$ & 1 &   & 1 &   & 1 &   & 1 &   & 1 &   & 1 & 2 & 1 &  & \tabularnewline
$N=14$ & 1 &   & 1 &   & 1 &   & 1 &   & 1 &   & 1 &   & 1 &  & 1\tabularnewline
\hline 
\hline
\end{tabular}
\end{table}

Apart from the scar tower, for $N=4L$, there are 2 additional $S^z=N-1$ states in the zero-energy sector, and for $N=4L+2$, the ferromagnetic state, which is not in the scar tower, is in the zero-energy sector.

Motivated by the fact that AKLT tower states also conserve total spin quantum number \cite{Mark2020}, we numerically compute the total spin of the the states in the zero-energy sector. The result is listed below:

\begin{table}[H]
\caption{Classification of zero-energy states with open boundary condition by the total spin number. The numbers in the table represent the number of states in each total $S$ sector.}
\centering{}
\setlength{\tabcolsep}{3mm}
\begin{tabular}{cccccccccccccccc}
\hline
\hline 
$S$ & 0 & 1 & 2 & 3 & 4 & 5 & 6 & 7 & 8 & 9 & 10 & 11 & 12 & 13 & 14\tabularnewline
\hline 
$N= 4$ & 1 &   & 1 & 2 & 1 &   &   &   &   &   &   &   &   &  &\tabularnewline
$N= 6$ & 1 &   & 1 &   & 1 &   & 1 &   &   &   &   &   &   &  & \tabularnewline
$N= 8$ & 1 &   & 1 &   & 1 &   & 1 & 2 & 1 &   &   &   &   &  & \tabularnewline
$N=10$ & 1 &   & 1 &   & 1 &   & 1 &   & 1 &   & 1 &   &   &  & \tabularnewline
$N=12$ & 1 &   & 1 &   & 1 &   & 1 &   & 1 &   & 1 & 2 & 1 &  & \tabularnewline
$N=14$ & 1 &   & 1 &   & 1 &   & 1 &   & 1 &   & 1 &   & 1 &  & 1\tabularnewline
\hline 
\hline
\end{tabular}
\end{table}

We see that the additional states in zero-energy space also conserve total spin quantum number.

\subsection{Scar Space of Symmetry-protected Topological States}
In this section, we discuss the possibility to construct a scar space which can be spanned by a set of matrix product states, which are all nontrivial symmetry-protected topological \cite{Gu2009,Chen2012,Lu2012} (SPT) states. Furthermore, we show that under such construction, we can have an SPT initial state, whose dynamical trajectory remains SPT.

Consider a unitary or anti-unitary group, $W$, represented by operator $\hat{D}(W)$, such that
\begin{eqnarray}
	[\hat{D}(w),\hat{D}(\tilde g)] |\psi_0\rangle = 0,\ 
	\forall w\in W, \tilde g \in \tilde G. 
	\label{spt-condition}
\end{eqnarray}
We further assume that the anchor state $|\psi_0\rangle$ is invariant under $\hat{D}(w)$. Since $\psi_0$ is short-range-entangled, it is either a trivial or nontrivial SPT, protected by $W$. Then we observe that $\hat{D}(\tilde{g})|\psi_0\rangle$ is the same SPT protected by $\hat{D}(W)$, because 
(i) $\hat{D}(\tilde{g})|\psi_0\rangle$ is invariant under $\hat{D}(W)$, and
(ii) $\hat{D}(\tilde{g})$ is onsite-unitary.

In this way, we can construct an SPT scar dynamics by choosing (i) a non-trivial SPT initial state $|\psi_0\rangle$ protected by $\hat{D}(W)$, and (ii) a quasi-symmetry which satisfies Eq.(\ref{spt-condition}). Then the scar Hamiltonian can be designed to have both the symmetry $W$ and the quasi-symmetry $\tilde G$.

We now show the anchor state and the quasi-symmetry we discussed in the main text as well as in \ref{apndx:MPSU1} meet these requirements. First, we show the anchor state (\ref{mps-anchor}) is a non-trivial SPT state protected by time-reversal symmetry. To see this, first observe that $\mathcal{T}$ can be realized as $d_L(\mathcal{T})=d_R(\mathcal{T})=i\sigma_y \mathcal{K}$ and $d(\mathcal{T})= \exp(iS_y\pi) \mathcal{K}$ where $\mathcal{K}$ is the complex conjugation. Because $d_{L,R}(\mathcal{T})$ is a nontrivial projective representation of time reversal ($d^2_{L,R}=-1$), $|\psi_0\rangle$ is nontrivial. 

Then we show that the U(1) quasi-symmetry action
\begin{eqnarray}
	\hat{D}(\tilde g(\theta)) = \bigotimes_{j=1}^{N}\exp(i \hat{S}_i^z \theta)
\end{eqnarray}
satisfies Eq.(\ref{spt-condition}). Note that $\hat{D}(\tilde{g})$ is a product of spin rotations. Since all spin rotations commute with time reversal, we automatically have Eq.(\ref{spt-condition}). We thus showed that every state $\hat{D}(\tilde{g})\psi_0$ is also a nontrivial SPT protected by time reversal.

Now we consider the construction of the scar Hamiltonian. The strategy is basically the same as in \ref{apndx:MPSU1}. By choosing $m=3$ cluster, we ended up having the same tower states and the local projector $\hat{P}$ as in \ref{apndx:MPSU1}. While now we wish the Hamiltonian also has time-reversal symmetry:
\begin{eqnarray}
	\hat{D}(\mathcal{T}) \hat{H} \hat{D}^{-1}(\mathcal{T}) = \hat{H},
\end{eqnarray}
where the time-reversal operation is represented by
\begin{equation}
	\hat{D}(\mathcal{T}) \equiv \bigotimes_{i=1}^{N} \exp{(i\pi \hat S^y_i)} \mathcal{K}.
\end{equation}
To make the Hamiltonian time-reversal symmetric, instead of inserting random real symmetry matrices into the general Hamiltonian
\begin{eqnarray}
	\hat{H} = \sum_j (1-\hat{P}_{[j,j+1]}) \hat{h}_{[j,j+1]} (1-\hat{P}_{[j,j+1]}),
\end{eqnarray}
we insert time-reversal-symmetric random hermitian matrices, defined by
\begin{eqnarray}
	\hat{h}_{[j,j+1]} 
	&=& \hat{D}(\mathcal{T}) \hat{m}_{[j,j+2]} \hat{D}^{-1}(\mathcal{T}) + \hat{m}_{[j,j+2]} \nonumber \\
	&=& \left[\otimes_{i=j}^{j+2}\exp{(i\pi \hat S^y_i)}\right] \hat{m}^{*}_{[j,j+2]} \left[\otimes_{i=j}^{j+2}\exp{(-i\pi \hat S^y_i)}\right] + \hat{m}_{[j,j+2]}, 
\end{eqnarray}
where $\hat{m}$'s are random complex Hermitian matrices, into the general form of Hamiltonian above.

We then numerically checked the Hamiltonian we designed (with size $N=9$) also obey Wigner-Dyson level statistics:

\begin{figure}[H]
\begin{centering}
\includegraphics[width=.5\linewidth]{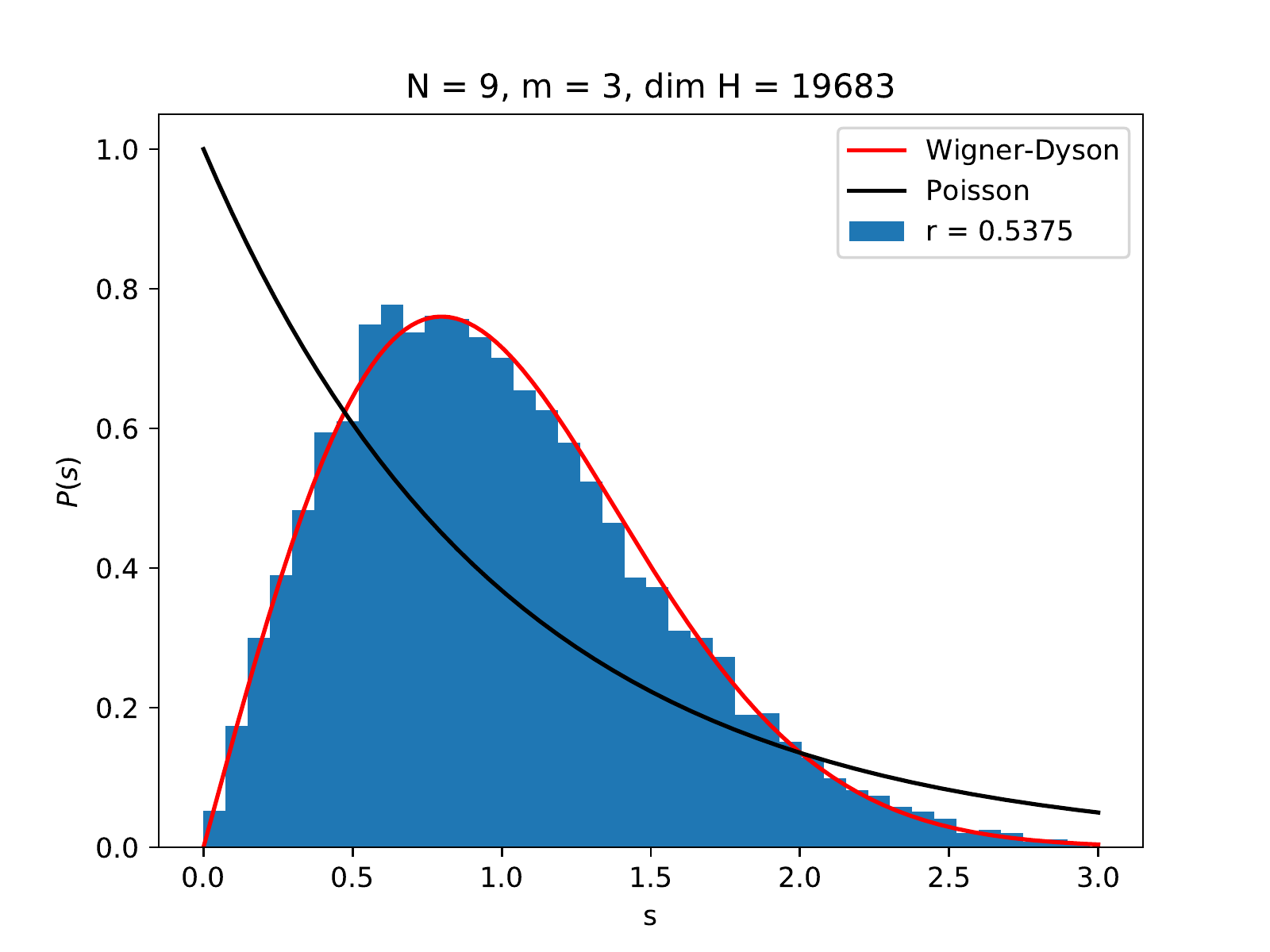}
\par\end{centering}
\caption{Distribution of many-body level spacing $s$ in the middle half of the spectrum of $\hat H$ for $N=9$ with periodic boundary condition. The Hilbert space dimension is 19683. The $r$-statistics is consistent with Wigner-Dyson GOE distribution for chaotic model \cite{PhysRevB.82.174411}.}
\end{figure}

In order to have SPT dynamics, we can add a new term to this Hamiltonian:
\begin{eqnarray}
	\hat{H}' = \hat{H} + \lambda \sum_{i=1}^{N} \hat{S}^z_j.
\end{eqnarray}
The initial state $|\psi_0\rangle$ will then have revival dynamics with period $\pi/\lambda$, and the whole trajectory are non-trivial SPT states.

\subsection{Dynamics of the Quantum Scar}
Here we discuss the dynamics of the states in the scar subspace. For a general scar tower with equally spacing energy
\begin{equation}
	E_n = E_0+f(n)\delta E,
\end{equation}
where $f(n)$ is an integer-valued function. A generic state in the tower-space, $|\psi\rangle=\sum_n c_n |\psi_n\rangle$, evolves as
\begin{equation}
	|\psi(t)\rangle=e^{-iE_0t}\sum_n c_n e^{-if(n)\delta{E}t}|\psi_n\rangle.
\end{equation}
Since $f(n)$ is an integer, we have
\begin{equation}
	|\psi(t+2\pi\delta{E}^{-1})\rangle=e^{-i 2\pi E_0 \delta E^{-1}}|\psi(t)\rangle.
\end{equation}
Two quantum states differing by an overall phase are identical.

The effect of generic random perturbation on the scar dynamics is previously studied \cite{Motrunich2020}. The result showed that a generic perturbation will eventually bring the scar state to thermalization, while the thermalization time can be very long if the perturbation is weak. In the following we sketch a simple proof of the statement, following the strategy in Ref.\onlinecite{Motrunich2020}.

Here we consider the one-dimensional Hamiltonian with a generic local perturbation:
\begin{equation}
	\hat H = \hat H_0 + \lambda \sum_i \hat V_i,
\end{equation}
where $\hat H_0$ is the scar Hamiltonian with equally spaced scar tower and support revival dynamics for special initial states. Each $\hat V_i$ only act non-trivially on the neighboring sites of site $i$, and the norm of $\hat V_i$ is bounded by a finite value $n_V$. 

Consider a local observable $\hat O_j$, which is defined on the neighboring sites of site $j$, and whose norm is bounded by another finite value $n_O$. The expectation value of $\hat O_j$ evolves as
\begin{eqnarray}
	\bar{O}_j(t) 
	&\equiv & \langle \psi(t)|\hat O_j |\psi(t)\rangle \nonumber \\
	&=& \langle \psi|e^{i \hat H t} \hat O_j e^{-i\hat H t}|\psi\rangle.
\end{eqnarray}
For unperturbed system, we showed that the expectation value will oscillate persistently and never converge to thermal value. The question is how fast the evolution of $\bar{O}_j(t)$ deviates from the oscillating trajectory. That is, we are seeking an upper bound for the quantity
\begin{eqnarray}
	\delta \bar{O}_j(t) 
	&\equiv& |\langle \psi|e^{i \hat H t} \hat O_j e^{-i\hat H t}|\psi\rangle - \langle \psi|e^{i \hat H_0 t} \hat O_j e^{-i\hat H_0 t}|\psi\rangle| \nonumber \\
	&=& |\langle \psi|e^{i \hat H t} \hat O_j e^{-i\hat H t} - e^{i \hat H_0 t} \hat O_j e^{-i\hat H_0 t}|\psi\rangle|.
\end{eqnarray}
The norm of an operator $\hat{M}$ is defined as:
\begin{eqnarray}
	\Vert \hat M \Vert \equiv \max_{\phi\in \mathcal H} |\langle \phi| \hat{M}(t) |\phi\rangle|.
\end{eqnarray}
We can use the operator's norm to bound the expectation value:
\begin{eqnarray}
	\delta \bar{O}_j(t) 
	&\le & \Vert e^{i \hat H t} \hat O_j e^{-i\hat H t} - e^{i \hat H_0 t} \hat O_j e^{-i\hat H_0 t} \Vert \nonumber \\
	&=& \Vert \hat O_j - e^{-i \hat H t}e^{i \hat H_0 t} \hat O_j e^{-i\hat H_0 t}e^{i\hat H t} \Vert \nonumber \\
	&\le & \int_0^t \left\Vert \frac{d}{ds}(\hat O_j - e^{-i \hat H s}e^{i \hat H_0 s} \hat O_j e^{-i\hat H_0 s}e^{i\hat H s}) \right\Vert ds \nonumber \\
	&=& \lambda \int_0^t \sum_i \Vert [\hat{V}_i, e^{i \hat H_0 s} \hat O_j e^{-i\hat H_0 s}] \Vert ds,
\end{eqnarray}
where in the second line, we use the fact that unitary transformations do not alter the norm, and in the third line, we use the inequality:
\begin{eqnarray}
	\Vert \hat M(t) \Vert 
	\le \Vert \hat M(0) \Vert + \int_0^t \left\Vert \frac{d}{ds}\hat{M}(s) \right\Vert ds
\end{eqnarray}
Then we use the result of Lieb-Robinson bound \cite{Lieb1972}:
\begin{equation}
	\Vert [\hat V_i,e^{i \hat H_0 s} \hat O_j e^{-i\hat H_0 s}] \Vert 
	\le 2\Vert\hat V_i\Vert \Vert\hat O_j\Vert \ e^{-a\left(|i-j|-v_{LR}s\right)},
\end{equation}
where $v_{LR}$ is the Lieb-Robinson velocity. The Lieb-Robinson bound told us that most contribution to the commutator is contained in $|i-j| \le v_{LR}t$, beyond which the contribution decays exponentially. The summation of this exponential tail is an $O(1)$ constant $c_0$, thus
\begin{eqnarray}
	\delta \bar{O}_j(t)
	&\le & \lambda \int_0^t \left( \sum_{|i-j| \le v_{LR}t} \Vert \hat V_i \Vert \cdot \Vert \hat O_j\Vert + c_0 \right)ds \nonumber \\
	&\le & \lambda \int_0^t \left(v_{LR} n_O n_V s + c_0\right) ds \nonumber \\
	&=& \frac{1}{2}\lambda v_{LR} n_O n_Vt^2 + c_0 \lambda t.
\end{eqnarray}
We thus know the thermalization time is at least $O(\lambda^{-1/2})$. A simple generalization of this proof shows that for higher dimensional system the thermalization time is at least $O(\lambda^{-1/(1+d)})$. 

However, note that the bound in general is far from tight. In the proof, we upper bound the deviation by the operator's norm, while actually the true expectation is taken oven scar state, which is a special non-thermal state in the Hilbert space. This means the exact expectation value can be much smaller than the norm the operator. Besides, the Lieb-Robinson bound in many cases is not a tight bound.

\end{appendix}

\end{document}